\documentclass{aa}  

\usepackage{graphicx}
\usepackage{txfonts}
\usepackage[utf8]{inputenc}
\usepackage{dcolumn}
\usepackage{supertabular}
\usepackage{rotating}
\usepackage{longtable}
\usepackage{lscape}
\usepackage{float}
\usepackage{subfig}
\usepackage{url}
\usepackage{color}
\usepackage{natbib}
\usepackage[dvipsnames]{xcolor}
\usepackage{ulem}
\usepackage{soul}
\usepackage[colorinlistoftodos]{todonotes}
\usepackage[version=4]{mhchem} 
\usepackage{supertabular}  
\usepackage[colorlinks=true,
            linkcolor=blue,
            urlcolor=blue,
            citecolor=blue]{hyperref}

\defcitealias{jacob2022_hygal}{Paper~I}
\defcitealias{Kim2023_hygal2}{Paper~II}

\newcommand{\kms}{\mbox{km s$^{-1}$}}
\newcolumntype{d}[1]{D{.}{\cdot}{#1}}
\newcolumntype{.}{D{.}{.}{-1}}

\begin{document}

   \title{HyGAL: Characterizing the Galactic ISM with observations of hydrides and other small molecules}

   \subtitle{III. The absorption lines of [O\,\textsc{i}], CH, and OH }
    \titlerunning{[OI\,\textsc{i}], CH, and OH absorption lines}
    
   \author{W.-J. Kim\inst{1,3} \and 
           A. M. Jacob\inst{2,3,1} \and 
           D. A. Neufeld\inst{2} \and 
           P. Schilke\inst{1} \and 
           H. Wiesemeyer\inst{3}  \and
           M. Gerin\inst{4} \and
           M. G. Wolfire\inst{5} \and
           V. Ossenkopf-Okada\inst{1} \and
           V. Valdivia\inst{6,7} \and
           E. Falgarone\inst{8} \and 
           D. C. Lis\inst{9}\and 
           S. Bialy\inst{10} \and      
           M. R. Rugel\inst{11} \and
          \'A. S\'anchez-Monge\inst{12,13} \and 
           M. Busch\inst{14} \and
           T. M\"{o}ller\inst{1} \and
           F. Wyrowski\inst{3} \and 
           D. Seifried\inst{1} \and 
           K. M. Menten\inst{3}\thanks{In memory of Karl M. Menten, who passed away suddenly before this work was completed. His inspiring advice and invaluable contributions will be sincerely missed.} \and
           A. Saintonge\inst{3}
           }

   \institute{I. Physikalisches Institut, Universit\"at zu Köln, Z\"ulpicher Str. 77, 50937 K\"oln, Germany\label{1}
        \and
        William H. Miller III Department of Physics \& Astronomy, Johns Hopkins University, Baltimore, MD 21218, USA\label{2}
        \and 
        Max-Planck-Institut f\"{u}r Radioastronomie, Auf dem H\"{u}gel 69, 53121 Bonn, Germany\label{3} \\
        \email{wjkim@mpifr-bonn.mpg.de}
        \and 
        LERMA, Observatoire de Paris, PSL Research University, CNRS, Sorbonne Universit\'{e}s, F-75014 Paris, France\label{4}
        \and 
        Department of Astronomy, University of Maryland, College Park, MD 20742-2421, USA\label{5}
        \and 
        Liant, 5 Av. Jean Jaur\`es, 34600, B\'edarieux, France \label{6}
        \and
        Department of Physics, Graduate School of Science, Nagoya University, Furo-cho, Chikusa-ku, Nagoya 464-8602, Japan\label{7}
        \and
        Laboratoire de Physique de l’ENS, ENS, Université PSL, CNRS, Sorbonne Université, Université de Paris, 24 rue Lhomond, F-75005 Paris, France\label{8}
        \and 
        Jet Propulsion Laboratory, California Institute of Technology, 4800 Oak Grove Drive, Pasadena, CA 91109, USA\label{9}
        \and Technion - Israel Institute of Technology, Haifa, 3200003 Israel\label{10}
        \and Deutsches Zentrum f\"{u}r Astrophysik (DZA), Theaterpassage 2, 02826 G\"{o}rlitz, Germany\label{11}
        \and Institut de Ci\`encies de l'Espai (ICE, CSIC), Campus UAB, Carrer de Can Magrans s/n, 08193, Bellaterra (Barcelona), Spain\label{12}
       \and Institut d'Estudis Espacials de Catalunya (IEEC), 08860 Castelldefels (Barcelona), Spain\label{13}
       \and National Radio Astronomy Observatory, 520 Edgemont Road, Charlottesville, VA 22903-2475, USA\label{14}
        }

   \date{Received ; accepted }

 
  \abstract
   {The HyGAL Stratospheric Observatory for Infrared Astronomy (SOFIA) legacy program aims at characterizing the interstellar medium in the Milky Way using hydrides, [C~{\sc ii}], and [O~{\sc i}] absorption lines with the 2.7~m SOFIA telescope toward twenty-five submillimeter-bright Galactic star-forming regions. As part of HyGAL, we investigated correlations among the known \ce{H2} tracers -- CH and OH from SOFIA observations, and \ce{HCO+} and CCH from ancillary absorption line data from ground-based telescopes. We also examined the abundance variation of neutral atomic oxygen, [O~{\sc i}], observed in absorption. CH, OH, \ce{HCO+}, and CCH all exhibit strong mutual correlations. OH in particular shows tight correlations with \ce{HCO+} and CCH, reflecting their linked chemical and physical pathways. Column density ratios among these \ce{H2} tracers are consistent with previous measurements in local diffuse clouds and remain uniform across Galactic environments and velocity intervals. The gas phase oxygen abundance relative to total hydrogen, $\langle X({\rm O})\rangle=N({\rm O})/N(\ce{H}_{\rm total})$, is $(3.09\pm0.64)\times10^{-4}$, slightly below the elemental solar value but consistent with the previous observations measuring gas-phase abundances. We also find that $N$(H~{\sc i}) decreases toward the regions where the molecular fraction exceeds $f_{\ce{H2}}^N \sim 0.5$, marking the onset of the molecular phase. While the atomic oxygen abundance remains roughly constant, the abundances of OH, \ce{HCO+}, and CCH increase with the molecular fraction. Gas traced by the \ce{HCO+} absorption corresponds to higher molecular fractions than that traced by H~{\sc i} and hydride ions, highlighting density variations in the diffuse-to-translucent ISM along different lines of sight.
   
   }

   \maketitle

%

\section{Introduction}
The transition from the atomic gas phase, in which hydrogen exists primarily as H~{\sc i}, to the molecular-dominated phase (\ce{H2}) is a key process in understanding the formation of molecular clouds from the diffuse interstellar medium (ISM) \citep[e.g.,][]{Snow2006, Sternberg2014,Bialy2016}. Direct measurements of \ce{H2} emission in diffuse gas are challenging due to the high energy above ground of the lowest rotational transition. However, observations of \ce{H2} are feasible through absorption in its ground electronic transitions near $\sim$1100\,$\AA$ (the Lyman and Werner bands of \ce{H2}) at far-ultraviolet (FUV) wavelengths, which are limited to lines of sight toward relatively nearby hot stars \citep{Savage1977_H2,Rachford2009} and high-latitude external galaxies \citep{Gillmon2006}. Consequently, such observations cannot easily probe diffuse molecular (defined as regions with the local fraction of gas-phase carbon in \ce{C+}, $f_{\ce{C+}}^n>0.5$ and the local fraction of molecular hydrogen, $f_{\ce{H2}}^n>0.1$; \citealt{Snow2006}) and translucent clouds (defined as those with $f_{\ce{C+}}^n>0.5$ but the local fraction of gas-phase carbon in CO, $f_{\ce{CO}}^n<0.9$) at greater distances, which exhibit higher extinction at optical and FUV wavelengths.

To overcome this limitation, hydrides -- particularly CH \citep[e.g.,][]{Sheffer2008}, \ce{H2O} \citep[e.g.,][]{Flagey2013_h2o}, and HF \citep[e.g.,][]{Neufeld2010_hf,Sonnentrucker2010_hf} -- have been found to be excellent proxies for \ce{H2} in diffuse and translucent clouds, as they display remarkably constant abundances relative to \ce{H2}. Among hydrides, CH observed at optical wavelengths around 430 nm and one of the first interstellar molecules detected \citep{Swings1937_CH}, has long been recognized as an excellent proxy for \ce{H2}. This is largely due to its tight linear correlation with \ce{H2} column density that is approximately $N$(CH)/$N$(\ce{H2})~$=3.5\times10^{-8}$ found toward nearby diffuse clouds by \cite{Sheffer2008}. Being one of the lightest hydrides, CH also plays a key role in initiating the formation of more complex hydrocarbon species. Subsequently, several observational studies have reported strong correlations between other small hydrocarbons such as CCH and \ce{c-C3H2}, observed at millimeter wavelengths, and CH transitions observed in the centimeter and submillimeter regimes \citep{Gerin2010_ch_cch,Lucas2000_cch_c3h2,Liszt2002}. These relationships suggest that the ratio between CH and these small hydrocarbons may provide an additional means of tracing \ce{H2} in translucent clouds. 

In addition, a significant fraction of \ce{H2} may reside in regions where CO abundance and column densities are extremely low to produce a detectable signal, the so-called CO-dark molecular gas \citep{Grenier2005_co-dark,Wolfire2010_CO_dark}, a phenomenon particularly prevalent in low-metallicity environments \citep[e.g.,][]{Bialy2015_OH_CH,Madden2020_CO_dark_low_metallicity}. Observational studies of OH in absorption and emission \citep[e.g.,][]{Li2018_OH_dark_moleculeargas,Nguyen2018_OH_dust,Busch2021_OH_COdark,Busch2024_OH_M31,Rugel2025_OH_HI} and \ce{HCO+} in absorption \citep[e.g.,][]{Hogerheijde1995_hcop_co-dark,Lucas1996_HCOp,Liszt2018_hcop_absorption, Liszt2019_co-dark_alma_hcop,Gerin2019_hcop_h2} have revealed OH and \ce{HCO+} as \ce{H2} proxies, tracing CO-dark molecular gas as well as CO-bright molecular gas in diffuse ISMs. These observational findings suggest that OH and \ce{HCO+} can serve as another potential tracer of CO-dark molecular hydrogen gas.

\begin{table*}[h!]
\begin{center}
\tiny
\caption{Summary of the SOFIA observations and detections.}
\label{tab:detection_data_info}
\begin{tabular}{l cccc c cccc c cccc}
\hline\hline
Source & \multicolumn{4}{c}{CH} & & \multicolumn{4}{c}{OH}  &  & \multicolumn{4}{c}{O\,{\textsc i}}\\\cline{2-5} \cline{7-10} \cline{12-15}
 & Flight & $D$ & $T_{\rm c}$ & $T_{\rm rms}$ & & Flight & $D$ & $T_{\rm c}$ & $T_{\rm rms}$ & & Flight & $D$ & $T_{\rm c}$ & $T_{\rm rms}$\\
  &  &  & [K] & [K] & &  &  & [K] & [K] & &  &  & [K] & [K]\\
\hline 
HGAL284.015$-$00.86 & OC9S & --$^{b}$ & --   & --   &&	OC9S & Y$^{a}$  & 2.4& 0.54 && OC9S & Y & 2.0& 0.16 \\ 
HGAL285.26$-$00.05  & OC9S & --$^{b}$ & --   & --   &&	OC9S & --$^{b}$ & --   & --    && OC9S & Y & 2.4& 0.20 \\ 
G291.579$-$00.431   & OC9S & --$^{b}$ & --   & --   &&	OC9S & Y        & 3.1& 0.42 && OC9S & Y & 0.8& 0.15 \\ 
IRAS~12326-6245$^*$     & OC9S & --$^{b}$ & --   & --   &&	OC9S & --$^{b}$ & --   & --    && OC9S & Y & 2.7& 0.20 \\ 
G327.3$-$00.60 	    & (2)  & Y        & 2.5& 0.03&&	(1)  & Y        & 3.6 & 0.24 && (2)  & Y &0.4& 0.03 \\ 
G328.307+0.423$^*$ 	    & OC9S & --$^{b}$ & --   & --   &&	OC9S & Y        & 5.2 & 0.45 && OC9S & Y & 2.5& 0.15 \\ 
IRAS~16060$-$5146$^*$   & (3)  & Y        &11.8& 0.12&&	OC9BC& Y        &10.9 & 0.43 && OC9BC& Y & 2.2& 0.20 \\ 
IRAS~16164$-$5046$^*$   & (1)  & Y        & 7.8& 0.08&&	OC9BC& Y        & 9.9& 0.41 && OC9BC& Y & 3.1& 0.15 \\ 
IRAS~16352$-$4721$^*$   & OC9S & --$^{b}$ & --   & --   &&	OC9S & Y$^{a}$  & 6.6& 0.81 && OC9S & Y & 1.8& 0.12 \\ 
IRAS~16547$-$4247$^*$   & OC9S & --$^{b}$ & --   & --   &&	OC9S & Y        & 5.2& 0.36 && OC9S & Y & 2.0& 0.10 \\ 
NGC~6334~I$^*$  	    & OC9BC& Y        &15.3& 0.12&&	OC9BC& Y        &17.3& 0.43 && OC9BC& Y & 9.0& 0.20 \\ 
G357.558$-$00.321       & OC9N & Y        & 1.6& 0.09&&	OC9S & --$^{b}$ & --   & --    && OC9N,OC9S& Y & 0.6& 0.23 \\
HGAL0.55$-$0.85$^*$     & OC9N & Y        & 4.2& 0.10&&	OC9S & --$^{b}$ & --   & --    && OC9N,OC9S& Y & 2.2& 0.20 \\
G09.62+0.19$^*$         & OC9N & Y        & 3.3& 0.12&&	OC9S & --$^{b}$ & --   & --    && OC9N & Y & 1.0& 0.14 \\
G10.47+0.03$^*$         & (4)     & Y &  4.8& 0.11 &&	OC9S, (1) & Y   & 6.9$^*$& 0.39$^*$ && OC9BC,OC9S& Y & 1.8 &  0.22 \\
G19.61$-$0.23$^*$       & OC9N & Y        & 2.8& 0.07&&	OC9S & --$^{b}$ & --   & --    && OC9N,OC9S& Y & 1.1 & 0.18 \\
G29.96$-$0.02$^*$       & OC9BC& Y        & 3.6& 0.08&&	OC9BC& Y        & 5.7& 0.53 && OC9BC& Y & 5.0& 0.39 \\
G31.41+0.31$^*$         & OC9S & --$^{b}$& --   & --   &&	OC9S & Y        & 1.9& 0.41  && OC9S & N & 0.04& 0.13  \\
W43~MM1$^*$             & OC9F & Y        & 2.1& 0.11&&	OC9F & Y        & 2.4& 0.60 && OC9F & N & 0.1& 0.26 \\
G32.80+0.19$^*$         & OC9BC& Y        & 3.8& 0.09&&	OC9BC& Y        & 4.1& 0.41 && OC9BC& Y & 2.6& 0.09 \\ 
G45.07+0.13$^*$         & OC9BC& Y        & 3.0& 0.07&&	OC9BC& Y        & 5.3& 0.38 && OC9BC& Y & 3.7& 0.33 \\
DR21$^*$                & OC8H & Y        & 1.7& 0.09&&   OC9F & Y        & 2.7& 0.63 && OC8H,OC9F& Y & 0.7& 0.19 \\
NGC~7538~IRS1$^*\dagger$       & OC8H & Y        & 3.3& 0.06&&   OC8H & Y        & 4.4& 0.42 && OC8H & Y & 3.0& 0.05 \\
W3~IRS5$^*\dagger$             & OC8H & Y        & 5.5& 0.06 &&   OC8H,OC9F& Y    & 8.6& 0.39 && OC8H,OC9F& Y & 6.9& 0.06 \\
W3(OH)$^*\dagger$              & OC8H & Y        & 5.3& 0.11 &&   OC8H,OC9F& Y    & 6.5& 0.27 && OC8H,OC9F& Y & 1.8& 0.06 \\
\hline 
\end{tabular}
\end{center}
\tablefoot{In Flight columns: (1) \cite{Wiesemeyer2016} (2) \cite{Jacob2019_oh_ch} (3) \cite{Jacob2020_ch} (4) \cite{Jacob2020_Arhp}. (${a}$) Observations done with 50-60\% observing time. (${b}$) No observing time was allocated. Column D exhibits the detections of absorption features against the observed background continuum sources. ``Y'' indicates a detection, while for the non-detections (N), tentative absorption features are present, but the confirmation of these features remains ambiguous due to poor signal-to-noise ratios. Asterisks (*) denote sightlines with absorption data for \ce{HCO+} and CCH, and others not in \citetalias{Kim2023_hygal2} are listed in Table\,\ref{tab:alma_archival}. ($\dagger$) The spectral lines and column densities for OH, CH, and [O\,\textsc{i}] in these sources are published as Figs.\,4, 5, and 6 of \citetalias{jacob2022_hygal}. }
\end{table*}

Atomic oxygen, together with atomic carbon, is a key element transitioning between atomic and molecular gas phases \citep[e.g.,][]{Hollenbach1991_diffuse_PDRs,Hollenbach1997_dense_PDR}, as it can persist in atomic form to of $A_{\rm v} \approx 8$ magnitude \citep[e.g.,][]{Hollenbach2009_O_av, Poglitsch1996_O}. Beyond $A_{\rm V} \approx 2.5$ mag, the attenuation of the FUV radiation field leads to a substantial drop in gas temperature and subsequently, a drop in the collisional excitation of the [O {\sc i}] fine-structure lines in these shielded regions \citep[e.g.,][]{Hollenbach1999_PDRs,Kaufman1999_OI,Roellig2007_PDR}. X-ray observations capable of probing oxygen in both gas and dust phases yield an average ISM oxygen abundance of $\langle X{\rm (O)} \rangle = (4.85 \pm 0.06) \times 10^{-4}$ (equivalent to 0.99 solar) \citep{Baumgartner2006_OI_x-ray}, consistent with the elemental solar value of $4.9\times10^{-4}$ \citep{Asplund2009_solar_abundances}. On the other hand, atomic oxygen fine-structure line [O\,{\textsc i}] absorption observations with the Stratospheric Observatory for Infrared Astronomy (SOFIA) 2.7~m telescope indicate relatively lower gas-phase oxygen abundances around $3\times10^{-4}$ \citep[e.g.,][]{Wiesemeyer2016,Lis2023}, suggesting oxygen depletion in the Galactic ISM. This depletion may be due to unidentified depleted oxygen (UDO), potentially involved in the formation of oxygen-bearing species, particularly OH and \ce{HCO+}, or locked into dust grains \citep[e.g.,][]{Baumgartner2006_OI_x-ray, Wiesemeyer2016,Lis2023}. Thus, understanding oxygen budget in the diffuse ISM is essential for interpreting \ce{H2} tracers across different environments.

This paper presents new submillimeter absorption observations of the rotational transitions of CH, OH, the [O\,\textsc{i}] 63\,$\mu$m fine-structure line, and of archival \ce{HCO+} and CCH absorption lines toward the 25 sightlines of the HyGAL SOFIA Legacy program. This work complements the previous studies of the HyGAL projects presented in \cite{jacob2022_hygal} (hereafter, \citetalias{jacob2022_hygal}) and \cite{Kim2023_hygal2} (hereafter, \citetalias{Kim2023_hygal2}), and it is the third paper (Paper III) in the series.
The observational procedures and data acquisition for the SOFIA CH, OH, and [O\,\textsc{i}] absorption lines, together with the data reduction process, are described in Sect.~\ref{sec:obs_data}. This section also includes information on previously published CH and OH data for sources not observed in the present work, as well as archival \ce{HCO+} and CCH absorption line data obtained with the Atacama Large Millimeter/submillimeter Array–Atacama Compact Array (ALMA–ACA).
The general observational results, including detections of absorption features, are presented in Sect.~\ref{sec:results}. The determination of column densities for all studied species is explained in Sect.~\ref{sec:xclass}. In Sect.~\ref{sec:discussion}, we discuss correlations among column densities of \ce{H2} proxy tracers and examine abundance variations relative to atomic and molecular hydrogen as a function of molecular gas fraction along different lines of sight. In addition, we also investigate oxygen abundances relative to atomic and molecular hydrogen. Finally, Sect.~\ref{sec:summary} summarizes the main results of this study.

\begin{table*}
    \caption{Information about the studied species, transitions, and used instruments.}
    \small
    \begin{center}
    \begin{tabular}{lccccrc rl }
    \hline \hline
    Species & \multicolumn{2}{c}{Transition} & Frequency & $A_{\text{u,l}}$ & \multicolumn{1}{c}{$E_{\text{u}}$} & Telescope  & HPBW & $\eta_{\rm MB}^{b}$\\ \cline{2-3}
    &  $J^{\prime} - J^{\prime\prime}$ & $F^{\prime} - F^{\prime\prime}$ & [GHz] & [s$^{-1}$] & \multicolumn{1}{c}{[K]} & \multicolumn{1}{c}{(Instrument)} & \multicolumn{1}{c}{[$^{\prime\prime}$]} & \\
    \hline 
    OH & $5/2- 3/2$ & $2^{-} - 2^{+}$ & 2514.2987(9) & 0.0137 & 120.75 & SOFIA 2.7~m (GREAT 4G4) & 11.2 & 0.52\\
    ~~$^2\Pi_{3/2}, N = 2-1$ & & $3^{-} - 2^{+}$ & 2514.3167(9)$^{a}$ & 0.1368\\
    & & $2^{-} - 1^{+}$ & 2514.3532(9) & 0.1231 \\
     CH &  $3/2-1/2$ & $1^{-}-1^{+}$ & 2006.74886(6) &  0.0111& 96.31 & SOFIA 2.7~m (GREAT LFA) & 13.5 & 0.66\\
      ~~$^2\Pi_{3/2}, N = 2-1$ & & $1^{-}-0^{+}$ & 2006.76258(6) & 0.0223 & \\
            & & $2^{-}-1^{+}$ & 2006.79906(6)$^{a}$ & 0.0335 &  \\
     O~{\sc i} $\quad {}^{3}P_{1}-{}^{3}P_{2}$& $2-1$ &  & 4744.7775(1) & $8.91\times10^{-5}$& 227.76 & SOFIA 2.7~m (GREAT HFA) & 6.3 & 0.65\\
     \hline
     \ce{HCO+} & $1-0$ &  & 89.1885247 & $4.19\times 10^{-5}$ & 4.28 & IRAM 30~m \& ALMA-ACA &  27.6$^{\dagger}$& 0.81 \\ 
     CCH $\quad \,\,N=1-0$ & $3/2-1/2$ &  $2-1$ & 87.316898$^a$ & $1.53\times 10^{-6}$ & 4.19 & IRAM 30~m \& ALMA-ACA& 28.2$^{\dagger}$ & 0.81 \\   
    \hline    
    \end{tabular}
     \end{center}
     \tablefoot{The transition information is sourced from the Cologne Database for Molecular Spectroscopy \citep[CDMS,][]{Meuller2005_cdms}, except CH frequencies from \citet{Truppe2014}. The uncertainty of the last digit of the frequency is indicated in parentheses. ($^{a}$) These indicate the hyperfine structure (hfs) transition that was used to set the velocity scale in the analysis. ($^{b}$) For the SOFIA data, the main-beam efficiencies for each channel (and, where applicable, each pixel) are determined for each flight series, and we list the typical values here. For \ce{HCO+} and CCH, the main-beam efficiencies ($\eta_{\rm MB}$) are those of the IRAM 30~m spectral data.  ($\dagger$) These are the beam sizes for IRAM 30 m observations of \ce{HCO+} and CCH, and are also used to extract molecular line spectra from ALMA-ACA archival data for the corresponding transitions.}
 \label{tab:spectroscopic_properties}
\end{table*}

\section{Observations and data reduction}
\label{sec:obs_data}

\subsection{SOFIA observations and data reduction}
\label{sec:sofia}
The HyGAL SOFIA Legacy Program targeted 25 continuum sources (listed in Table~\ref{tab:hygal_complete_source_list}), selected based on their continuum levels at 160\,$\mu$m from the Hi-GAL survey (Photodetector Array Camera and Spectrometer onboard the Herschel Space Observatory; \citealt{Molinari2010_Higal}). The selection criteria required $F_{\rm 160\,\mu m} > 2000$\,Jy for sources within the inner Galaxy (Galactocentric distance, $R_{\rm GAL} < 8$\,kpc) and $F_{\rm 160\,\mu m} > 1000$\,Jy for those in the outer Galaxy ($R_{\rm GAL} > 8$\,kpc); further details are provided in Sect. 3.1 of \citetalias{jacob2022_hygal}. The HyGAL observations were carried out with the 2.7\,m SOFIA telescope over multiple campaigns during Cycles~8 and~9 (Project ID:~08\_0038; PIs: D.~A.~Neufeld and P.~Schilke). High–spectral resolution data were obtained using the German REceiver for Astronomy at Terahertz Frequencies (GREAT) heterodyne instrument. The receiver configurations included the 4GREAT module \citep{Duran2021} and the High-Frequency Array (HFA) and Low-Frequency Array (LFA) channels of upGREAT \citep{Risacher2018}. The CH ($\nu = 2.006$\,THz) and [O\,\textsc{i}] ($\nu = 4.744$\,THz; 63\,$\mu$m) lines were observed with the LFA and HFA, respectively, while the OH ($\nu = 2.514$\,THz) transition was observed with 4GREAT. Details of the GREAT observing campaigns, including CH, OH, and [O\,\textsc{i}] observations and detections of absorption features, are provided in Table~\ref{tab:detection_data_info}. The observed transitions and corresponding receiver configurations are summarized in Table~\ref{tab:spectroscopic_properties}.

\begin{figure}[h!]
    \centering
    \includegraphics[width=0.34\textwidth]{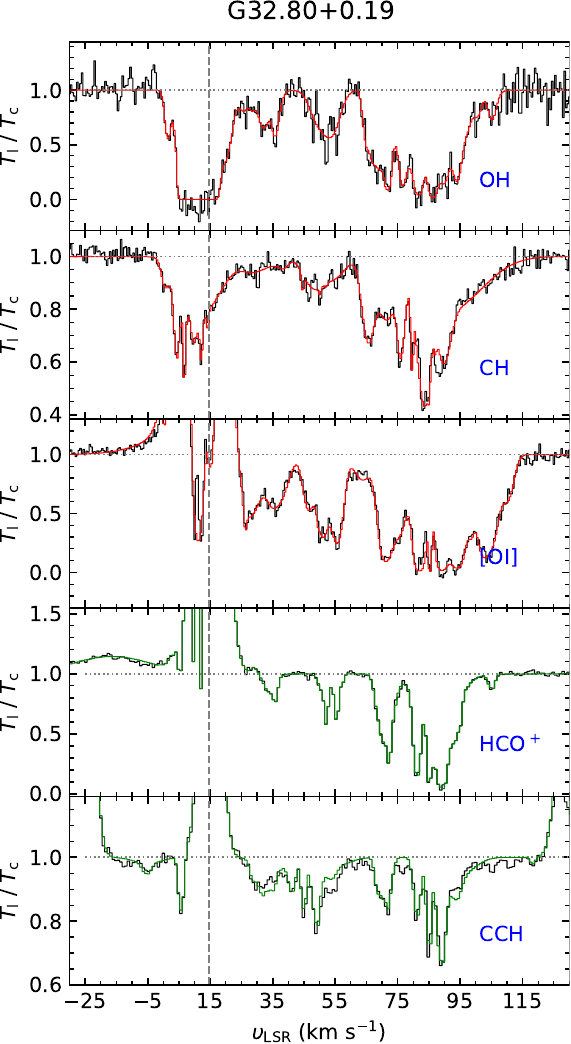}
    \caption{Observed spectra (black) of OH, CH, [O\,\textsc{i}], \ce{HCO+}, and CCH with overlaid XCLASS modeled spectra (red and green) toward G32.80$+$0.19 as an example. The vertical dashed lines indicate the systemic velocity ($\varv_{\rm sys}$) of the background continuum source.}
    \label{fig:spectra1}
\end{figure}

The observations were performed in double-beam switch mode, employing a chopping technique to minimize atmospheric and instrumental fluctuations. A modified version of the Fast Fourier Transform Spectrometer (FFTS) backend \citep{Klein2012} was used to provide 4\,GHz-wide spectral coverage with high spectral resolution. In the subsequent analysis, only the central pixel of each upGREAT array was used, as the continuum flux density decreases significantly in the outer pixels. These outer pixels are separated from the central pixel, and from each other, by approximately 32$''$ in the HFA and 14$''$ in the LFA (see Table\,\ref{tab:spectroscopic_properties} for beam sizes). The native spectral channel spacings correspond to velocity resolutions of $\Delta\upsilon = 0.036$, 0.029, and 0.2\,\kms\ for the CH, OH, and [O\,\textsc{i}] lines, respectively.

Prior to data analysis, all spectra were calibrated using the Kölner Observatorium für SubMillimeter Astronomie (KOSMA) calibrate task  \citep{Guan2012} to correct for residual atmospheric opacity at the flight altitude. The calibrated data were subsequently converted to the main-beam temperature scale using the beam efficiencies listed in Table~\ref{tab:spectroscopic_properties}. All standard data reduction procedures—including baseline subtraction, inspection for residual instrumental features, and spectral smoothing—were performed using the Continuum and Line Analysis Single-dish Software (CLASS\footnote{\url{https://www.iram.fr/IRAMFR/GILDAS/doc/html/class-html/class.html}}) within the Grenoble Image and Line Data Analysis Software (GILDAS) package \citep{Pety2005_gildas}. In several cases, the achieved sensitivity fell short of the expected values. Therefore, to improve the signal-to-noise ratio and to produce a uniform data set in terms of velocity resolution, all spectra were smoothed to a velocity spacing of $\Delta\upsilon = 0.5$\,\kms. These smoothing procedures represent an optimal balance between radiometric sensitivity and spectral resolution, as required to resolve the narrowest line profiles in our data. They also ensure that the individual hyperfine structure (hfs) components are spectrally resolved, with separations between hfs transitions ranging from 2.0 to 5.4\,\kms. An exception was made for the [O\,\textsc{i}] spectrum toward NGC~6334~I, which already exhibits excellent sensitivity at a native resolution of 0.2\,\kms. Additional smoothing in this case would artificially merge intrinsically narrow spectral features and was therefore not applied. The average $T_{\rm rms}$ values are 0.45\,K for OH, 0.09\,K for CH, and 0.16\,K for [O\,\textsc{i}].

\subsection{Ancillary data and data reduction}
\label{sec:ancillary}
We used available SOFIA archival spectral data for OH (G327.3$-$00.60 and G10.47$+$0.03), CH (G327.3$-$00.60 and G10.47$+$0.03), and [O\,\textsc{i}] (G327.3$-$00.60) from the literature listed in Table\,\ref{tab:detection_data_info}. The OH data toward G10.47$+$0.03, taken during the campaign  OC9S, was significantly affected by total-power instability, and thus, we discarded it and used the archival data instead. We also note that the CH spectrum of G10.47$+$0.03 and the [O\,\textsc{i}] spectrum of G327.3$-$0.6 have velocity channel widths of 0.73\,\kms\ and of 1.0\,\kms, respectively, slightly broader than those of the HyGAL SOFIA spectra smoothed to 0.5\,\kms.

In addition, we utilized millimeter absorption line data for \ce{HCO+} and CCH, published in Sect.\,3.2 of \citetalias{Kim2023_hygal2} for the northern hemisphere sources and the ALMA-ACA archival data (Project ID: 2019.1.00685.S and PI: Tie Liu) for the southern hemisphere sources. Those sources with available \ce{HCO+} and CCH data are indicated in Table\,\ref{tab:detection_data_info}. For the ALMA-ACA archival data, we averaged spectra over the IRAM 30~m beam size, centered on the coordinates of the HyGAL sources (coordinates in Table\,\ref{tab:hygal_complete_source_list}) to ensure consistent area coverage for comparison with the IRAM 30~m data. Table\,\ref{tab:alma_archival} lists the information about the ALMA archival data toward the southern hemisphere sources, not included in \citetalias{Kim2023_hygal2}, and the determined continuum levels as well as noise levels ($T_{\rm rms}$). Furthermore, to enhance signal-to-noise ratios and achieve uniform spectral resolution consistent with SOFIA data, we smoothed the ALMA-ACA spectra to a channel width of $\Delta\upsilon=$~0.5\,\kms\ from the original $\Delta\upsilon$ of 0.1 and 0.2\,\kms.

To investigate abundance variations of the target species in diffuse and translucent clouds, it is essential to determine the molecular hydrogen fraction, which requires measuring the column density of neutral atomic hydrogen via H\,\textsc{i} 21\,cm absorption observations along with measuring the column density of molecular hydrogen. To this end, we conducted H\,\textsc{i} 21\,cm line observations with the Karl G. Jansky Very Large Array (JVLA) for the northern sources (Project IDs: 20A-519 and 21A-287; PI: M.~R.~Rugel) and analyzed archival data from the Australia Telescope Compact Array (ATCA) for the southern sources. The derivation of H\,\textsc{i} column densities for the HyGAL sources will be presented in detail in Rugel et al. (in prep.), and the methodology is described in Sect.\,3.1 of \citet{Rugel2025_OH_HI}.

\section{Results}\label{sec:results}
Figure~\ref{fig:spectra1} presents an example of the observed absorption spectra normalized by the continuum brightness temperature ($T_{\rm l}/T_{\rm c}$) of OH, CH, [O\,\textsc{i}], \ce{HCO+}, and CCH (shown as black profiles) toward G32.80$+$0.19 as a function of local standard of rest velocity ($\upsilon_{\rm LSR}$). As seen in Fig.\,\ref{fig:spectra1} as an example, OH ($^2{\Pi}_{3/2}$, $J=5/2-3/2$; top panel) suffers more than CH ($^2\Pi_{3/2}$, $J=3/2-1/2$, second panel from the top) from saturated absorption. The red and green curves indicate the modeled profiles obtained using the eXtended CASA Line Analysis Software Suite (XCLASS; version~1.4.2\footnote{\url{https://xclass.astro.uni-koeln.de}}
; \citealt{moeller2017}). The details of the spectral fitting procedure are provided in Sect.~\ref{sec:xclass}. For the \ce{HCO+} and CCH transitions, the modeled spectra for sources observed with the IRAM~30\,m telescope are adopted from \citetalias{Kim2023_hygal2}, while the corresponding ALMA–ACA spectra are presented and analyzed in Sect.~\ref{sec:xclass}.  The spectral line plots toward all other sources are presented in Figs.\,\ref{appendix:spectra1} and \ref{appendix:spectra2}.

Toward all the observed lines of sight, we detected absorption features of CH, OH, and [O\,\textsc{i}], except in the [O\,\textsc{i}] spectra toward W43~MM1 and G31.41$+$0.31. At the central pixel position of W43 MM1, there are no [O\,\textsc{i}] features observed in either emission or absorption. However, weak [O\,\textsc{i}] emissions are detected at sightline velocities between $+$102\,\kms\ and $+$106\,\kms\ in the outer pixels, specifically at offsets ($\Delta x$, $\Delta y$) = ($-7.5''$, $-12.4''$) and ($+6.9''$, $-13.0''$). By considering the systemic velocity of W43~MM1, which is 97.8\,\kms, we can associate these emission components with this complex. For G31.41$+$0.31, there is a good [O\,\textsc{i}] detection in emission. The lack of detected absorption features toward W43~MM1 and G31.41$+$0.31 may be attributed to the weak continuum levels of the sources at 4.744\,THz, with $T_{\rm{cont}}=$ 0.1\,K and 0.04\,K, respectively.

\begin{figure}[t!]
    \centering
    \includegraphics[width=0.34\textwidth]{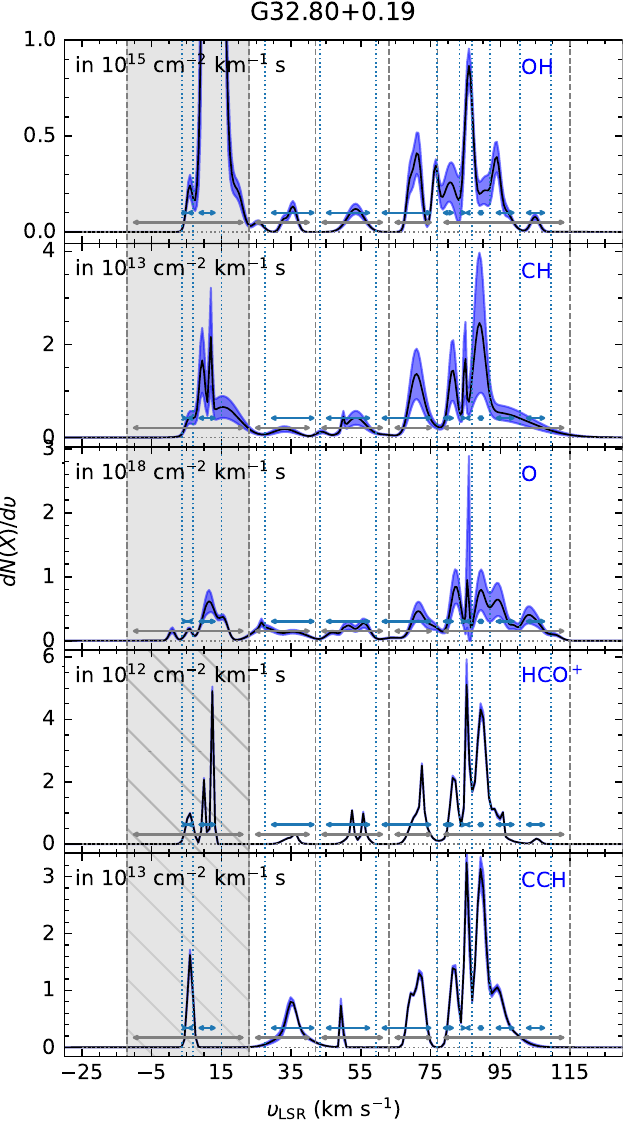}
    \caption{Channel-wise column density spectra, $dN(X)/d\varv$, as a function of velocity for OH, CH, O, \ce{HCO+}, and CCH from top to bottom, toward G32.80$+$0.19 as an example. Column density units are shown in the upper left corner. The gray areas in the column density plots represent the velocities of background sources, which also show emission features within this velocity range. Blue-shaded areas indicate fitting uncertainties. The vertical gray dashed lines, as well as gray arrows, mark velocity intervals of VI(A), while the light-blue dotted lines, as well as blue arrows, denote intervals of VI(M). Column density plots for non-displayed sources are shown in Fig. \ref{appendix:Ncol}.}
    \label{fig:Ncol_spec}
\end{figure}

\subsection{Determination of column densities with XCLASS}
\label{sec:xclass}
XCLASS is used to model spectral lines studied here by solving the 1-dimensional radiative transfer equation. Here, we assumed an isothermal source. To model and generate synthetic spectra fitted to the observational spectra, we used the \texttt{myXCLASSFit} function. This function computes a synthetic spectrum by fitting the observational spectral lines with Gaussian profiles, thereby ensuring that opacity effects on the line shapes are properly considered. The details of the fitting procedures using XCLASS are described in Sect.\,4.2 of \citetalias{jacob2022_hygal} and Sect.\,4.1 of \citetalias{Kim2023_hygal2}. 

The initial parameter set for fitting each absorption or emission line component includes the excitation temperature ($T_{\rm ex}$), the total column density ($N_{\rm tot}$), the full-width half maximum line width ($\Delta \varv_{\rm FWHM}$), and the velocity offset ($\varv_{\rm off}$) from the systemic velocity. The column density per velocity interval, $dN/d\varv$ for each velocity channel, $i$, is related to the optical depth, $\tau_i$ by
\begin{equation}
\begin{aligned}
\label{eq:optically_thin}
\left(\frac{dN}{d\upsilon}\right)_i = \frac{8\pi\nu^{3}}{c^{3}} \frac{Q(T_{\rm ex})}{g_{u}A_{ul}} \frac{{\rm exp} ({E_{u}}/{k_{\rm B}T_{\rm ex}} )}{{\rm exp}(h\nu/{k_{\rm B}T_{\rm ex}}) -1} \tau_i,
\end{aligned}
\end{equation}
where $E_{\rm u}$ and $g_{\rm u}$ are the energy and degeneracy of the upper state of a given transition at a rest frequency, $\nu$. $A_{\rm ul}$ is the Einstein coefficient for spontaneous emission. $Q$($T_{\rm ex}$) and $k_{\rm B}$ are the partition function and Boltzmann constant, respectively. The fractional populations across the hfs split sub-levels, rotational levels, and the fine-structure line of [O~{\sc i}] are set by radiative equilibrium with the cosmic microwave background temperature of 2.73\,K as an excitation temperature ($T_{\rm ex}$) (for [O~{\sc i}], a lower limit). XCLASS describes a layered structure of gas along a line of sight. In this study, absorption features are considered a ``foreground'' absorbing gas layer, while emission features are regarded as an emitting background ``core'' layer.

The best-fit parameters (i.e., $N_{\rm tot}$, $\Delta\upsilon_{\rm FWHM}$, and $\upsilon_{\rm off}$) are obtained by minimizing the $\chi^2$ value using the Levenberg-Marquardt algorithm. In addition, in the fitting procedure, all the hfs splitting transitions of CH and OH, as well as CCH, are taken into account by fitting all the transitions of a selected molecule simultaneously, as shown in Fig.\,\ref{fig:spectra1}. Thus, the outcome of the fitting deconvolves the hfs of these molecular transitions. For some lines of sight, the absorption lines of \ce{HCO+} may be affected by contamination from broad outflow components appearing in emission (e.g., see Sect.\,3.2 of \citetalias{Kim2023_hygal2} for a detailed description of broad emission features related to outflows). In particular, the spectral lines of [O~{\sc i}], \ce{HCO+}, and CCH show absorption features blended with emission features at the velocities of the background sources (vertical dashed lines in Figs.\ref{fig:spectra1}, \ref{appendix:spectra1}, and \ref{appendix:spectra2}). To accurately determine column densities of absorption components, we simultaneously fitted the absorption and emission features of [O~{\sc i}], \ce{HCO+}, and CCH. In the case of OH, the absorption features from the sightlines are significantly stronger than any potential emission. Therefore, we fitted only the OH absorption features. Some strong OH absorption features having broad profiles with high-velocity wings indicate outflow signatures. These features can be easily distinguished by comparing CH or \ce{HCO+} absorption or emission line profiles. They could also be identified by comparing 1835/1837\,GHz data from excited OH appearing in emission. Because these lines (i.e., $^2\Pi_{1/2}, J = 3/2 - 1/2$) connect higher energy levels ($E_{\rm u}=270.14$\,K for $F=2^+-1^-$) than the $^2\Pi_{3/2}, J=5/2 - 3/2$ lines studied here ($E_{\rm u}=120.75$\,K for $F=2^- - 2^+$ and connecting to the ground state), they appear in emission \citep{Csengeri012_oh_outflow,Csengeri2022_OH_outflow}. The blue-shifted OH absorption toward NGC~6334~I shows a clear excess absorption at velocities below about $-$20\,\kms, compared to the CH profile. On the other hand, toward other sources, there is no significant velocity excess in OH absorption, suggesting that OH outflows are not detected. However, we cannot fully exclude potential contamination due to outflows in the OH absorptions without a detailed analysis by comparing 1835/1837\,GHz data of OH transitions. In addition, the uncertainties in column densities of all species studied here are determined by sampling a posterior distribution of the optical depths ($\tau=-ln(T_{\rm l}/T_{\rm c}$)). The details of the error estimation are also described in  Sect.\,4.2 of  \citetalias{jacob2022_hygal} and Sect.\,4.1 of \citetalias{Kim2023_hygal2}.

Figure\,\ref{fig:Ncol_spec} shows, as an example, velocity-specific column density profiles, $dN({\rm X})/d\varv$, of OH, CH, O, \ce{HCO+}, and CCH from top to bottom panels. Reconstructing emission profiles without good knowledge of the source structure is problematic. Thus, fitting absorption features blended by complicated emission line profiles introduces significant uncertainties in the column density values. For \ce{HCO+} and CCH, absorption features affected by self-absorption or heavily blended with emission features (shown in the hatched gray areas) are excluded due to unreliable column density measurements, especially toward DR21, G29.96$-$0.02, G10.47$+$0.036, and G09.62$+$0.19. In addition, we used column density profiles and absorption spectra of atomic (A) and molecular (M) tracers to define two distinct velocity intervals (VIs). The velocity interval VI(A) is based on broad or continuous absorption profiles of H~{\sc i} (\citealt{Rugel2025_OH_HI} and Rugel et al. in prep.) and \ce{OH+} (e.g., \citetalias{jacob2022_hygal} and Jacob et al. in prep.) as these species are sensitive to the gas components of diffuse clouds with lower molecular gas fraction (e.g., for \ce{OH+}, molecular gas fractions between $1.5\times10^{-3} - 3\times10^{-2}$; \citealt{Jacob2020_Arhp}). Another set of velocity intervals VI(M) is based on \ce{HCO+} absorption features as explained in Sects.\,5.2 and 5.3 of \citetalias{Kim2023_hygal2}, indicating to trace translucent clouds or denser parts of diffuse clouds. The velocity range, VI(A), generally encompasses broader intervals compared to VI(M). This is because the \ce{HCO+} absorption components are not as broad or continuous as those of H~{\sc i}  or \ce{OH+}. As a result, the column densities integrated over VI(A) account for more diffuse or translucent components of molecular clouds along the line of sight, compared with column densities over VI(M). In Figs.\,\ref{fig:Ncol_spec} and \ref{appendix:Ncol}, the VI(A) and VI(M) are indicated by gray and blue horizontal arrows, respectively. 

We adopt the column density ratio (3.5$\times10^{-8}$; \citealt{Sheffer2008}) of CH and \ce{H2} to determine the \ce{H2} column density and molecular gas fraction. The validity of the linear relationship between CH and \ce{H2} is established across various Galactic scales \citep[e.g.,][]{Federman1982,Sheffer2008,Weselak2019}, but the deviations and significant scatter remain to be understood. The \ce{H2} column densities toward all the available HyGAL lines of sight are derived by adopting the abundance of CH relative to \ce{H2} toward the most of sources and the median ratio value of $N$(OH)/$N$(CH), which is 3.29$\pm$0.11, measured in this work, toward the sources without CH data. Unlike the other \ce{H2} proxy in diffuse and translucent clouds, HF ($J=1-0$; e.g., \citealt{Sonnentrucker2015_HF}) absorption lines, in addition, OH, and CH absorption lines suffer less from saturation. The determined $N$(\ce{H2}) column densities over VI(A) are listed in Table\,\ref{tab:column_density_tb1}.

\begin{figure}
    \centering
    \includegraphics[width=0.38\textwidth]{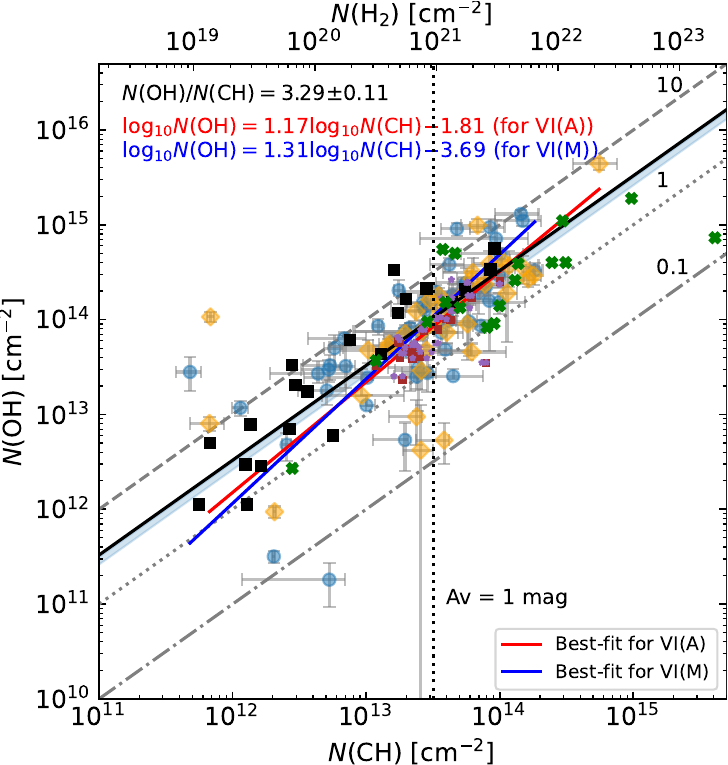}
    \caption{Column densities of OH versus CH integrated over VI(A) (orange diamonds) and VI(M) (blue circles). Previous observations from literature are overlapped: \cite{Wiesemeyer2016} -- green x, \cite{Weselak2019} -- brown squares, \cite{Jacob2019_oh_ch} -- black squares, and \cite{Mookerjea2016_ch_oh} -- purple asterisks. The upper x-axis is for $N$(\ce{H2}) values derived by the empirical CH-\ce{H2} relationship ($N$(CH)/$N$(\ce{H2}) $=3.5\times10^{-8}$; \citealt{Sheffer2008}). The red and blue solid lines represent the best-fit results for VI(A) and VI(M) data points, respectively. The black solid line represents the median value of the $N$(OH)/$N$(CH) ratios. Gray diagonal lines indicate different ratios (0.1, 1, and 10) of column densities. The blue shaded area indicates the ratio ranges found in previous studies \citep[i.e.,][]{Wiesemeyer2016, Jacob2019_oh_ch, Rawlins2023, Liszt2002}. The vertical dotted line represents a visual extinction ($A_{\rm v}$) of 1 magnitude. }
    \label{fig:Nch_Noh}
\end{figure}
\begin{table*}[h!]
\centering
\small
\caption{Linear fitting results for the relationships.}\label{tab:linearfit}
    \begin{tabular}{c c c c c c }
    \hline \hline
       Velocity intervals  & $x$ & $y$ & a$\pm\Delta a$ & log$_{10}(b)\pm \Delta$log$_{10}(b)$ & $r_{\rm s}$ \\
       \hline
       VI(A)  & $N$(CH) & $N$(OH) &1.17$\pm$0.19 & $-$1.81$\pm$2.62 & $0.79^{+0.10}_{-0.17}$\\
       VI(M) & $N$(CH) & $N$(OH) & 1.31$\pm$0.20 & $-$3.69$\pm$2.74 & $0.80^{+0.08}_{-0.12}$ \\
       \hline
       VI(A)   & $N$(CH) & $N$(\ce{HCO+}) & 1.36$\pm$0.17 & $-$5.94$\pm$2.36 & $0.80^{+0.10}_{-0.16}$ \\
       VI(M) & $N$(CH) & $N$(\ce{HCO+})& 1.12$\pm$0.10 & $-$2.64$\pm$1.39  & $0.83^{+0.06}_{-0.10}$ \\
       \hline
       VI(A)   & $N$(CH) & $N$(CCH) & 1.05$\pm$0.11 & $-$0.77$\pm$1.54 & $0.79^{+0.11}_{-0.21}$ \\
       VI(M) & $N$(CH) & $N$(CCH) & 1.14$\pm$0.11 & $-$1.94$\pm$1.47 & $0.72^{+0.11}_{-0.17}$\\
       \hline
       VI(A)   & $N$(OH) & $N$(\ce{HCO+}) & 1.04$\pm$0.14 & $-$2.13$\pm$1.92 & $0.84^{+0.08}_{-0.14}$ \\
       VI(M) & $N$(OH) & $N$(\ce{HCO+}) & 1.02$\pm$0.09 & $-$1.94$\pm$1.29 & $0.87^{+0.05}_{-0.08}$ \\
       \hline
       VI(A) & $N$(OH) & $N$(CCH) & 1.11$\pm$0.13 & $-$2.17$\pm$1.86 & $0.56^{+0.23}_{-0.35}$\\
       VI(M) & $N$(OH) & $N$(CCH) & 1.03$\pm$0.09 & $-$0.94$\pm$1.23 & $0.67^{+0.14}_{-0.19}$\\
       \hline
    \end{tabular}
    \tablefoot{The fits discussed in Sect.\,\ref{sec:discussion} were performed in log space, where log$_{10}(y) = a$ log$_{10}(x) +$ log$_{10}(b)$ is for the relationship of $y=bx^a$. All the Spearman correlation coefficients ($r_{\rm s}$) have $p-$values much smaller than 3$\sigma$. }
\end{table*}
\begin{figure*}
    \centering
    \includegraphics[width=0.38\textwidth]{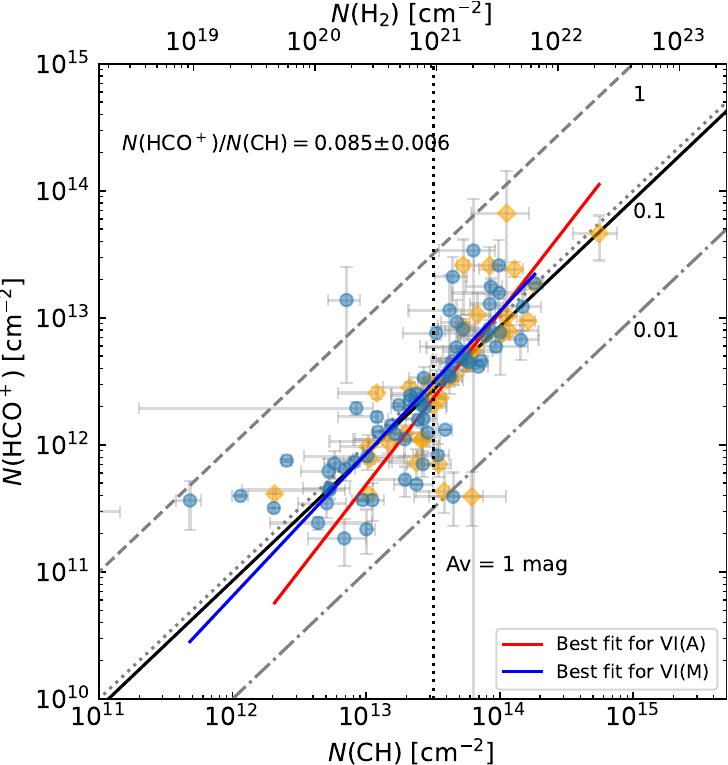}
    \hspace{0.8cm}
    \includegraphics[width=0.38\textwidth]{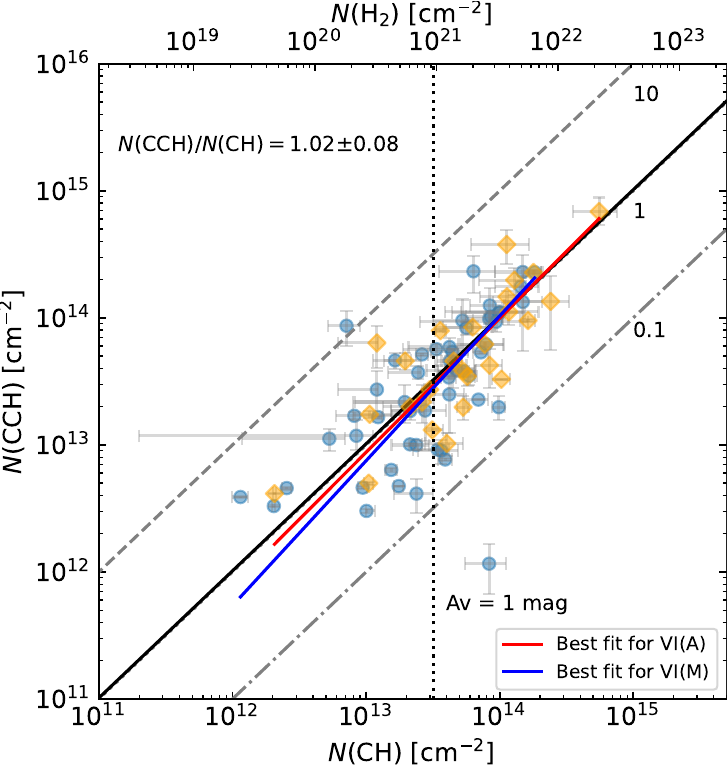}
    \caption{Velocity-integrated column density comparisons of CH versus \ce{HCO+} (left panel) and CCH (right panel). The markers indicating data points determined in this study are the same as in Fig.\,\ref{fig:Nch_Noh}. The blue and red solid lines are the best-fit results of our data. The black solid lines represent the median values of the column density ratios of CH and \ce{HCO+} (left panel) and those of CH and CCH (right panel). Gray diagonal lines indicate different ratios (in the left panel, 1, 0.1, and 0.01, and in the right panel, 10, 1, and 0.1) of column densities. The upper axis of both plots represents the column density of \ce{H2}.}
    \label{fig:Nch_N3mm}
\end{figure*}

\section{Analysis and Discussion}
\label{sec:discussion}
\subsection{Column density correlations among \ce{H2} proxy tracers}

Figure\,\ref{fig:Nch_Noh} shows the column density comparison between CH and OH determined over the velocity intervals of VI(A) and VI(M) from this work and those taken from previous absorption line studies \cite[i.e.,][]{ Mookerjea2016_ch_oh, Wiesemeyer2016, Weselak2019, Jacob2019_oh_ch}. Due to the ambiguity in the optical depths of the absorption components spanning the associated background source velocity range, we considered only the foreground absorption components in all analyses, including the bisector regression fitting \citep{Akritas1996_becs}, considering uncertainties in both variables as independent, and calculating Spearman correlation coefficients ($r_{\rm s}$). All the data points of foreground components with different velocity intervals (blue circles for VI(M) and orange diamonds for VI(A)) are well distributed between the unity slope (dotted line) and a ratio of 10 (dashed line), and they also show a good agreement with the previous measurements. We obtained excellent correlations toward both velocity interval samples, showing $r_{\rm s}$ of 0.79$_{-0.17}^{+0.10}$ (for VI(A)) and 0.80$_{-0.12}^{+0.08}$ (for VI(M)), both with $p$-value $\ll$\,3$\sigma$. The best-fits with the bisector regression of the two data sets are log$_{10}$$N$(OH)~$=$~1.17$\pm$0.19~log$_{10}N$(CH)~$-1.81$$\pm2.62$ (red solid line) for VI(A) data points and log$_{10} N$(OH)~$=$~1.31$\pm$0.20~log$_{10}N$(CH)~$-3.69\pm2.74$ (blue solid line) for VI(M) data points. The fitting results for both samples are very similar and slightly steeper than the unit slope (gray dotted line). The marginally steeper slopes are likely influenced by the limited number of data points and the considerable scatter observed at lower column densities ($N(\mathrm{CH}) \lesssim 3\times10^{12}$ cm$^{-2}$ and $N(\mathrm{OH}) \lesssim 10^{13}$ cm$^{-2}$).

\begin{figure*}[t]
    \centering
    \includegraphics[width=0.38\textwidth]{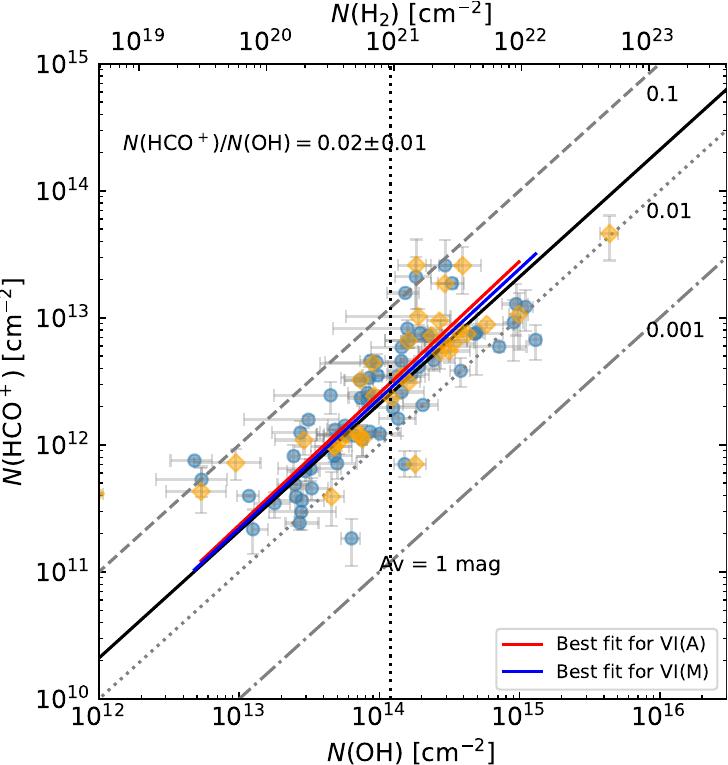}
    \hspace{0.8cm}
    \includegraphics[width=0.38\textwidth]{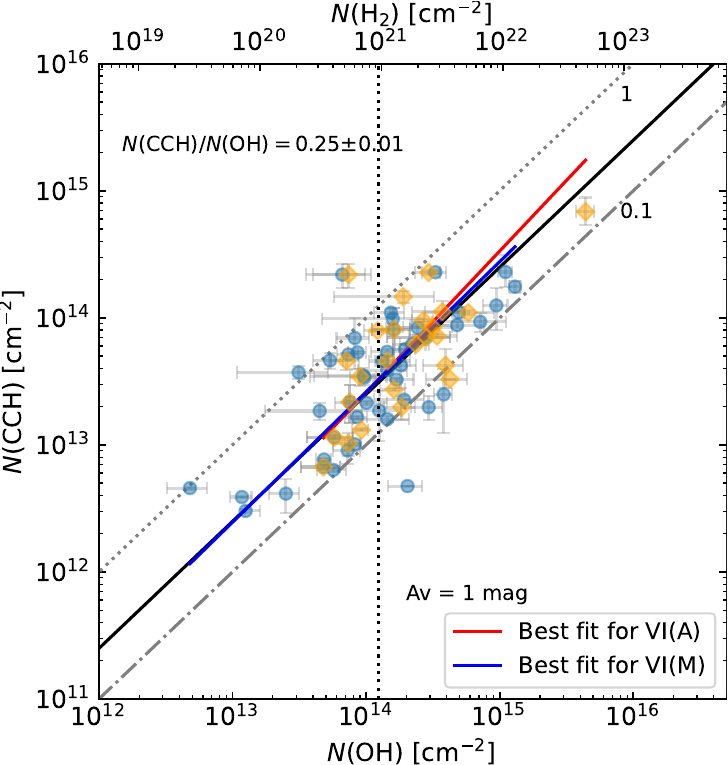}
    \caption{Velocity-integrated column densities of CCH and \ce{HCO+} versus OH. The markers indicating data points determined in this study are the same as in Fig.\,\ref{fig:Nch_Noh}. The blue and red solid lines are the best-fit results of our data. The black solid lines represent the median values of the column density ratios of OH and \ce{HCO+} (left panel) and those of OH and CCH (right panel). Gray diagonal lines indicate different ratios (in the left panel, 0.1, 0.01, and 0.001, and in the right panel, 1 and 0.1) of column densities, and the upper axis of both plots represents the column density of \ce{H2}.}
    \label{fig:Noh_N3mm}
\end{figure*}

The median $N$(OH)/$N$(CH) ratio determined from the data points defined over VI(A) and VI(M) is 3.29$\pm$0.11, in good agreement with the value of 3.14$\pm$0.49 reported in previous studies \citep[e.g.,][]{Wiesemeyer2016, Jacob2019_oh_ch}, which were based on the same CH and OH transitions observed toward different Galactic diffuse clouds. Interestingly, the ratio derived in this work is also similar with those obtained from near-UV absorption measurements (CH at 3137.576 and 3143.150\,$\AA$, and OH at 3078.443 and 3081.664\,$\AA$; 2.61$\pm$0.19; \citealt{Rawlins2023}) and from radio observations (CH at 3.335\,GHz, \citealt{Liszt2002}; OH at 1.665 and 1.667\,GHz, \citealp{Lucas1996_HCOp}), which yielded a ratio of 3.0$\pm$0.9 toward local diffuse clouds. Despite these diffuse and translucent clouds being located at different Galactocentric radii (see Table
\,\ref{tab:hygal_complete_source_list}) and positions within the Milky Way (MW), the close agreement across all studies suggests that CH and OH maintain a remarkably consistent abundance relationship across a wide range of Galactic environments.

By comparing our sightline components with those from \citet{Jacob2019_oh_ch} (shown as black squares), we find that our data points are predominantly concentrated in the region where $N$(CH)~$>4\times10^{12}$\,cm$^{-2}$ and $N$(OH)~$>10^{13}$\,cm$^{-2}$. In contrast, most of the lower column density points from the work of \citet{Jacob2019_oh_ch} correspond to the sightline toward AGAL330.954$-$00.18, where the low CH and OH column densities are associated with inter-arm velocity components. The sensitivities achieved in our observations are insufficient to detect absorption features with optical depths below $\sim$0.1. Moreover, the lack of data points at low column densities is likely related to the source selection. Any weaker inter-arm absorption components are likely blended with stronger spiral-arm features, resulting in single, broader absorption profiles.

To examine the relationships between CH, OH, and other commonly used \ce{H2} proxies, we compared CH and OH with \ce{HCO+} and CCH. Figure\,\ref{fig:Nch_N3mm} presents the column density comparisons of $N$(CH) with $N$(\ce{HCO+}) (left panel) and $N$(\ce{CCH}) (right panel). The results of the bisector regression fittings and the corresponding $r_{\rm s}$ are listed in Table\,\ref{tab:linearfit}. Overall, both CCH and \ce{HCO+} show strong correlations with CH, with $r_{\rm s}~=0.83-0.72$, comparable to those found between CH and OH. The median column density ratios, $N$(\ce{HCO+})/$N$(CH) $=0.085\pm0.006$ and $N$(\ce{CCH})/$N$(CH) $=1.02\pm0.08$, are indicated by black solid lines in Fig.\,\ref{fig:Nch_N3mm}. Although CH shows a slightly weaker correlation with CCH than with \ce{HCO+}, their median ratio and the best-fits (red: log$_{10}N$(CCH) $=1.05\pm0.11$\,log$_{10}N$(CH)$-0.77\pm1.54$ and blue: log$_{10}N$(CCH) $=1.14\pm0.11$\,log$_{10}N$(CH)$-1.94\pm1.47$) suggests an approximately 1:1 relationship across all sightlines, consistent with previous studies \citep[e.g., $N$(CCH)/$N$(CH)~$=$~0.6–1.2;][]{Gerin2010_ch_cch}. This implies that CCH and CH likely share similar formation and destruction pathways, which involves \ce{C2H2+} during formation and \ce{C+} during destruction (for details, see the simplified chemical networks for CH and CCH in (A) of Fig.\,\ref{fig:chemical_networks}), as well as likely originate from the same or similar gas \citep[e.g.,][]{Gerin2010_ch_cch}. Despite these correlations, the $N$(\ce{HCO+})/$N$(CH) and $N$(\ce{CCH})/$N$(CH) ratios span roughly one order of magnitude. A similar scatter between CH and these species has been observed in Galactic diffuse and translucent clouds on sightlines toward extragalactic targets \citep[e.g.,][]{Liszt2002}.  This suggests that the diffuse molecular clouds in the Galactic disk and in the off-disk environment are likely similar, implying the environmental impact on their chemistry is not significantly different. However, based on time-dependent chemical network calculations from the magnetohydrodynamics (MHD) simulations conducted by \cite{Komichi2026_CH_OH_HCOp}, the column densities of \ce{HCO+} and \ce{H2} exhibit a non-linear relationship. In contrast, the column densities of OH and CH show a linear dependence on \ce{H2}. The authors suggest that the non-linear relationship between \ce{HCO+} and \ce{H2} can be attributed to the formation pathways of \ce{HCO+} (e.g., the sketch (B) of Fig.\,\ref{fig:chemical_networks}), which depend on CO abundances (i.e., CO-rich and CO-poor -- CO-dark regions). This may explain the non-linear relationship observed between CH and \ce{HCO+} at an $A_{\rm v}$ of approximately 1 mag (as shown in the left panel of Fig.\,\ref{fig:Nch_N3mm}) found in this study. This highlights the importance of investigating whether these variations are linked to the time-dependent evolution of diffuse molecular clouds in CO-dark and CO-rich environments.

The OH column densities show a tight correlation with those of \ce{HCO+} (left panel of Fig.\,\ref{fig:Noh_N3mm}), where such tight correlations were also found in the previous observations \citep[e.g.,][]{Liszt1996_OH}. The scatter in this relationship is smaller than that between CH and \ce{HCO+}, with ratios confined within roughly one order of magnitude. For the OH and CCH comparison, the number of data points, especially for lower CCH column densities ($<10^{13}$\,cm$^{-2}$), is insufficient for strong statistical conclusions, yielding $r_{\rm s}=0.56$ ($p=0.004$) for VI(A) and $r_{\rm s}=0.67$ ($p \ll 3\sigma$) for the VI(M) intervals. However, the linear best-fits show a clear relationship in column density between OH and CCH (right panel in Fig.\,\ref{fig:Noh_N3mm}). The median column density ratios are $N$(\ce{HCO+})/$N$(OH)$=0.02\pm0.01$ and $N$(\ce{CCH})/$N$(OH)$=0.25\pm0.01$. According to the results of principal component analysis (PCA) in Sects.\,4.2 and 5.1 of \citetalias{Kim2023_hygal2} toward two specific sightlines (see also Figs.\,13, 14, 15, and 16 of \citetalias{Kim2023_hygal2}), OH correlates slightly better with \ce{HCO+} and CCH, though CH and OH remain closely associated, as in this work, showing a strong correlation between CH and OH toward more sightlines.  In fact, \ce{HCO+} and OH are closely linked via one of the primary chemical pathways: OH + \ce{C+} $\rightarrow$ \ce{CO+} $+$ H followed by \ce{CO+} $+$ \ce{H2} $\rightarrow$ \ce{HCO+} $+$ H \citep{Godard2009_TDR,Gerin2019} (see the simplified chemical networks for OH and \ce{HCO+} in (B) of Fig.\,\ref{fig:chemical_networks}). In addition, \cite{Jacob2020_Arhp} pointed out that CH and OH abundances peak at a molecular gas fraction, $f_{\ce{H2}}$, higher than 0.1. We further discuss these abundance correlations across different molecular fractions in Sect.\,\ref{sec:abundance_fraction}. Additionally, no significant differences are found between the two velocity-interval samples. The consistent column density ratios for CH, OH, \ce{HCO+}, and CCH across all intervals suggest similar chemical conditions in diffuse and translucent molecular clouds, consistent with those observed in the diffuse ISM of the solar neighborhood and along extragalactic sightlines.

\begin{figure}[t!]
    \centering
    \includegraphics[width=0.38\textwidth]{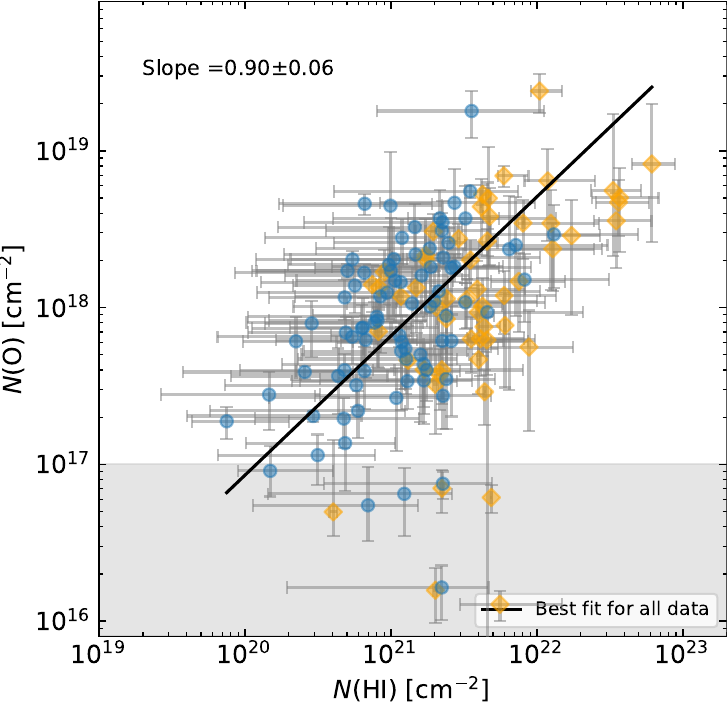}
    \vskip 0.2cm
    \includegraphics[width=0.39\textwidth]{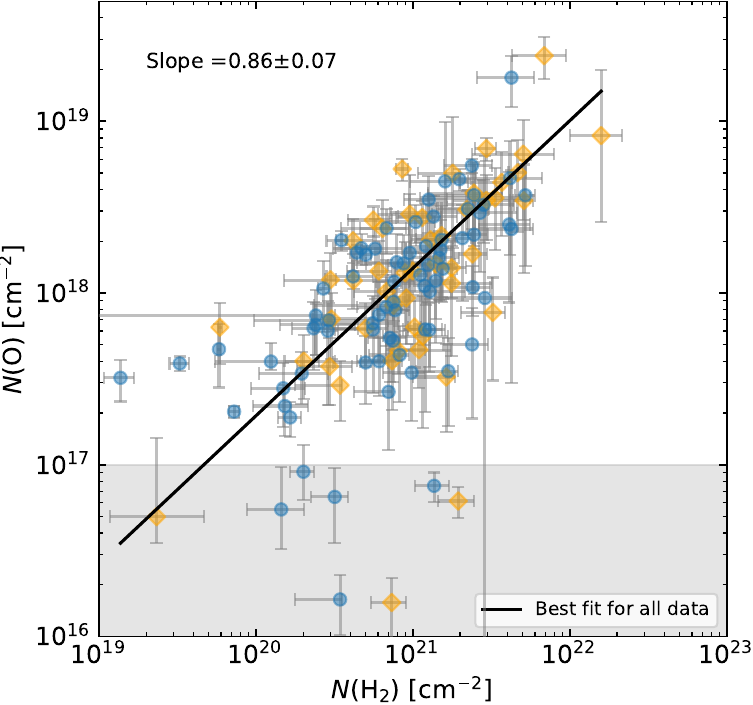}
    \caption{Atomic oxygen column density as functions of atomic hydrogen column density (upper) and molecular hydrogen column density (lower). In the plots, we only consider foreground velocity components due to the ambiguity caused by the potential contamination of emission lines. In addition, we neglect saturated absorption features ($e^{-\tau} \approx 0$). The \ce{H2} column density is primarily determined by the CH column density. However, for cases without CH observations, the OH column density is used after converting it to the CH column density using the OH/CH ratio. The black solid lines represent the best regression fitting results. The $N$(O) below $10^{17}$~cm$^{-2}$ is close to the general noise limit, and thus any data points with $N$(O) $<$ $10^{17}$~cm$^{-2}$ in the gray shaded area are excluded in determination of the oxygen abundances relative to $N$(H~{\sc i}) and $N$(\ce{H2}) and the best fitting.}
    \label{fig:No_Nhi_Nh2}
\end{figure}

\begin{figure}[]
    \centering
    \includegraphics[width=0.38\textwidth]{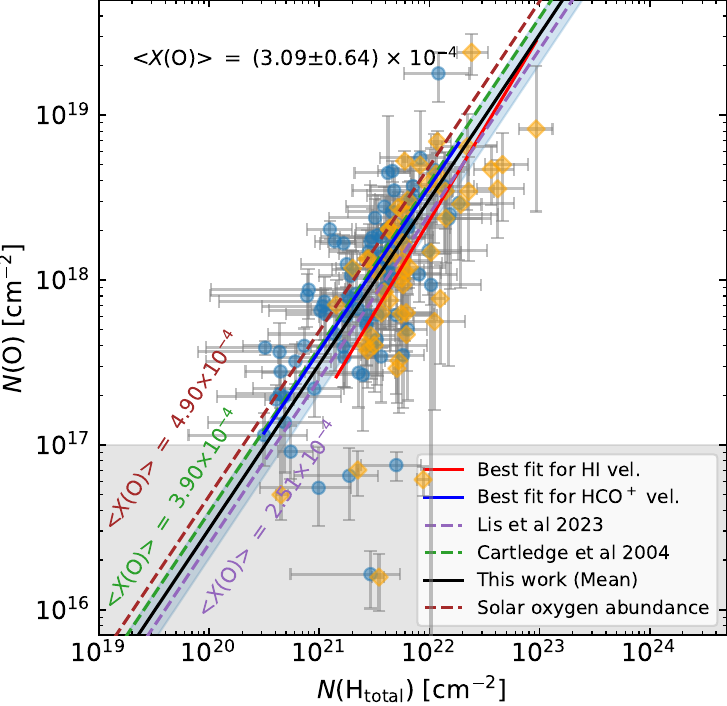}
    \caption{Atomic oxygen column density as a function of total hydrogen nucleus column density. The markers indicating data points and the gray shaded area are the same as in Fig.\,\ref{fig:No_Nhi_Nh2}. The blue shaded area represents the range of the standard deviation of the atomic oxygen abundances. }
    \label{fig:No_Ntotal}
\end{figure}

\subsection{Neutral atomic oxygen column density}
As discussed in Sect.\,5.1 of \citetalias{Kim2023_hygal2}, the neutral atomic oxygen column densities toward W3(OH) and W3~IRS5, both associated with local arm clouds, show strong correlations with CH and OH, with 91–94\% correlations. At the same time, a considerable correlation was observed between O and ionized species such as \ce{C+}, \ce{H2O+}, \ce{OH+}, and \ce{ArH+} (correlations of 60–95\%; Figs.\, 15 and 16 of \citetalias{Kim2023_hygal2}), which trace diffuse clouds with molecular gas fractions, $f_{\ce{H2}}<0.1$ \citep[e.g.,][]{Snow2006,Indriolo2015_H2Op_OHp_fmol}. These results agreed with the fact that oxygen remains predominantly atomic in deeper regions up to $A_{\rm v}\sim8$\,mag, across diffuse and translucent cloud regimes.

To further investigate the association of neutral oxygen with atomic and molecular hydrogen, we compared $N$(O) with $N$(H~\textsc{i}) and $N$(\ce{H2}) for all HyGAL sources. As shown in the upper panel of Fig.\,\ref{fig:No_Nhi_Nh2}, $N$(O) increases with $N$(H~{\sc i}), following a log–log slope of 0.90$\pm$0.06 from the bisector regression fitting and a Spearman correlation coefficient of $r_{\rm s}=0.59$ ($p \ll 3\sigma$). A spatial separation in $N$(H~{\sc i}) ($\sim 2\times10^{21}$\,cm$^{-2}$) is also visible between the H~{\sc i} and the \ce{HCO+} velocity intervals (blue circles for VI(M) and orange diamonds for VI(A); see upper panel of Fig.\,\ref{fig:No_Nhi_Nh2}). In contrast, the comparison with $N$(\ce{H2}) (lower panel) shows no such separation, suggesting that part of the observed oxygen arises in low-density diffuse clouds where atomic hydrogen dominates. Nonetheless, the moderate correlation between $N$(O) and $N$(\ce{H2}) ($r_{\rm s}=0.67$, $p \ll 3\sigma$) with a fitted slope of 0.86$\pm$0.07 implies that neutral atomic oxygen is also associated with diffuse and translucent regions containing significant amounts of molecular hydrogen.

Figure\,\ref{fig:No_Ntotal} presents the relationship between $N$(O) and the total hydrogen column density, $N$({H$_{\rm total})=N$(H~{\sc i})$+ 2N$(\ce{H2}). The black solid line indicates the median abundance of atomic gas-phase oxygen relative to total hydrogen, $\langle X$(O)$\rangle$ = $N$(O)/$N$({H$_{\rm total}$), with a value of $(3.09 \pm 0.64) \times 10^{-4}$, which is in agreement with abundances of 3.1 to 3.5 toward W31C, G34.26 and W49N \citep{Wiesemeyer2016}.  The median abundances for the VI(A) and VI(M) samples are $2.46 \times 10^{-4}$ and $3.73 \times 10^{-4}$, respectively. The median abundance aligns well with the best-fit lines (red; log$_{10}$ $N$(O) $=$ 1.12$\pm$.13 log$_{10}$ $N$(H$_{\rm total}$) $-$ 6.33$\pm$2.77  and blue; log$_{10}$ $N$(O) = 1.00$\pm$0.07 log$_{10}$ $N$(H$_{\rm total}$) $-$ 3.33$\pm$1.47) derived for the two data sets (orange diamond and blue circle markers), although the fit for the VI(A) sample deviates slightly.

The oxygen abundances derived here are consistent with previous absorption measurements, such as $2.51\times10^{-4}$ \citep{Lis2023} toward the Srg B2 sightline crossing several lines of sight inter- and spiral-arms and $3.90\times10^{-4}$ \citep{Cartledge2004} toward O- and B-type stars in the Galactic disk, confirming the robustness of our results. This agreement suggests that oxygen abundances across the MW are relatively uniform, with environmental variations reflected in the scatter around the median. Along the Sgr~B2 sightline, \citet{Lis2023} reported variations of $0.65\times10^{-4}$ in oxygen abundance among different Galactic plane regions and spiral arms, consistent with the dispersion seen in the HyGAL sources. Within the uncertainties, all VI(A) data points lie below the elemental solar oxygen abundance of $4.9\times10^{-4}$ \citep{Asplund2009_solar_abundances}, while several VI(M) measurements extend above it. These higher values likely reflect uncertainties in $N$(H$_{\rm total}$) and $N$(O), but they remain plausible given that elevated oxygen abundances have also been reported in previous studies \citep[e.g., $(5.75\pm0.4)\times10^{-4}$;][]{Przybilla2008A_oxygen_abundance}. In addition, there are notable increases in oxygen abundances within the spiral arms, as indicated by the significant depths of neutral oxygen absorption shown in Figs.\,1 and 2 of \cite{Lis2023} and in our measurements. Our results also support this observation, showing that, at a given total hydrogen column density, the VI(M) data points associated with the strong neutral oxygen absorption depths generally exhibit higher neutral oxygen column densities than the VI(A) data points, which integrate over larger volumes of diffuse envelope gas. The existence of inhomogeneous oxygen abundances within diffuse clouds is evident, even when accounting for measurement uncertainties. This variation is not limited to differences across the Galactic disk radius, as the measurements from the different velocity intervals, VI(M) and VI(A), reveal significant discrepancies in these abundances, underlining the complexity of the interstellar medium.

\begin{figure}[t!]
    \centering
    \includegraphics[width=0.38
    \textwidth]{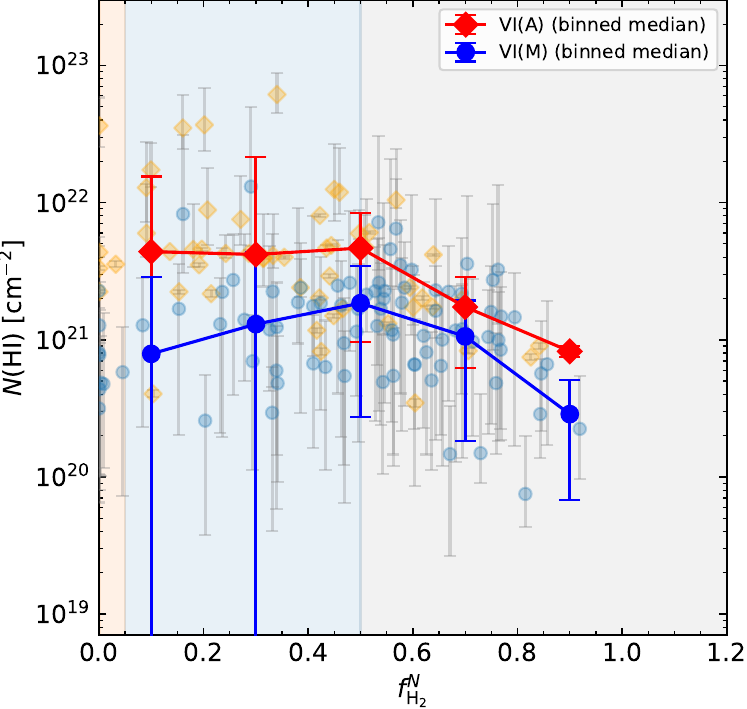}
    \caption{Neutral atomic hydrogen column density, $N$(H~{\sc i}), versus molecular gas fraction. The color codes of data points are the same as Fig.\,\ref{fig:Nch_Noh}. The different shaded regions indicate the atomic phase ($f_{\rm \ce{H2}}^{N}<0.05$, orange shaded area), transition phase ($0.05 < f_{\ce{H2}}^N <0.5$, blue shaded area), and molecular phase ($f_{\ce{H2}}^N > 0.5$, gray shaded area). The red and blue markers are the averaged $N$(H~{\sc i}) over a bin of 0.1 $f_{\ce{H2}}^N$, for the VI(A) and VI(M) samples. Their error bars are the standard deviation of the averaged data points. }
    \label{fig:Nh2_Nhi}
\end{figure}
\begin{figure}[t!]
    \centering
    \includegraphics[width=0.38\textwidth]{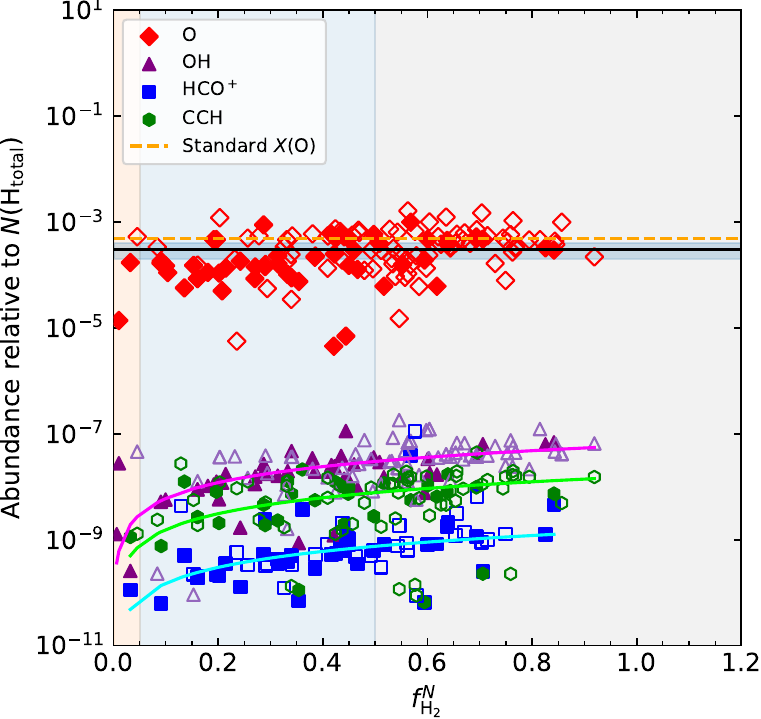}
    \includegraphics[width=0.38\textwidth]{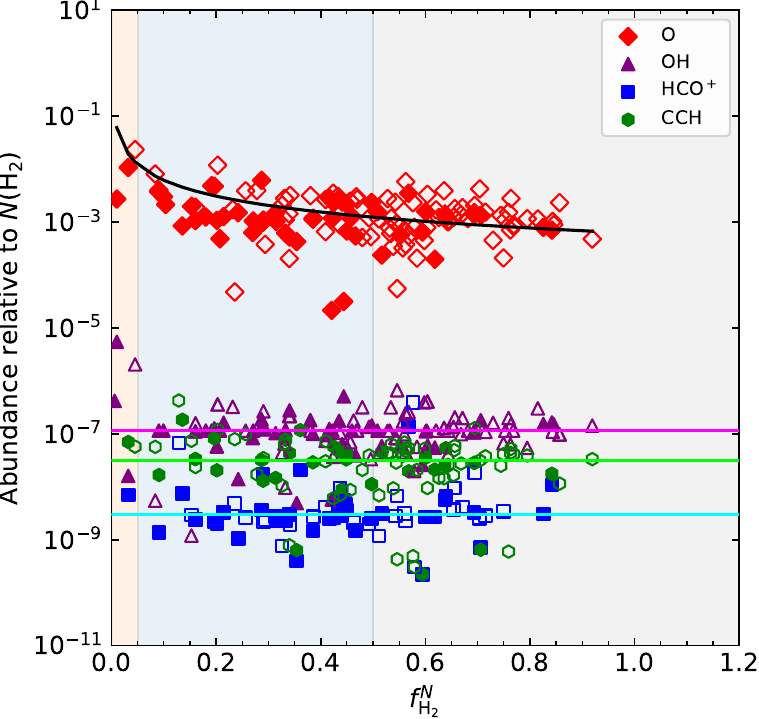}
    \caption{Fractional abundances relative to total hydrogen (upper panel) and to $N$(\ce{H2}) (lower panel) as a function of molecular gas fraction ($f_{\ce{H2}}^{N}$), for different species compared. The different colored shaded regions are the same as Fig.\,\ref{fig:Nh2_Nhi} (atomic phase in orange, transition phase in blue, and molecular phase in gray). The different-colored markers indicate different species. The filled markers are data points for VI(A) intervals, and the open markers are for VI(M) intervals. }
    \label{fig:x_fmol}
\end{figure}
\subsection{Variation in abundances of oxygen, OH, \ce{HCO+}, and CCH according to molecular gas fractions}
\label{sec:abundance_fraction}
The classification of interstellar clouds is determined by the relative fractions of \ce{H2}, \ce{C+}, and CO in the gas-phase as well as by their visual extinction, gas density, and temperature \citep{Snow2006}. In particular, within diffuse molecular clouds, the fraction of \ce{H2} varies substantially, reflecting the transition from H to \ce{H2}, while the electron density remains nearly constant \citep{Neufeld2005,Snow2006}. Understanding this H-to-\ce{H2} transition in diffuse molecular cloud regions is therefore crucial for characterizing the chemical and physical evolution of the ISM. To examine how the abundances of the species studied here vary with the molecular gas fraction in the Galactic ISM, given that their column densities exhibit different dependencies on the atomic and molecular hydrogen components, we estimated the molecular gas fraction, $f^N_{\ce{H2}}$, to be the line of sight fraction, expressed as a ratio of column densities using the following relation \citep[e.g.,][]{Savage1977_H2, Bohlin1978_paper2}, 
\begin{equation}\label{eq:fh2_1}
    f^{N}_{\ce{H2}} = \frac{2N(\ce{H2})}{N({\rm H~{I}}) + 2N(\ce{H2})}. 
\end{equation}

Figure \ref{fig:Nh2_Nhi} presents a comparison between $N$(H {\sc i}) and $f_{\ce{H2}}^N$. The samples defined over the VI(A) intervals exhibit systematically higher $N$(H {\sc i}) values than those integrated over the VI(M) intervals, particularly in regions where the molecular gas fraction is below 0.5, corresponding to the atomic and transition phases (highlighted by the orange and blue shaded regions). These results might suggest that the broader VI(A) intervals encompass more diffuse material, whereas most of the \ce{H2} gas is concentrated within the narrower velocity ranges that are traced by \ce{HCO+} absorption lines. However, within the uncertainties of the column density measurements, $N$(H {\sc i}) remains approximately constant across both samples up to $f_{\ce{H2}}^N \sim 0.5$. 
In contrast, within the molecular gas regime ($f_{\ce{H2}}^N > 0.5$; gray-shaded region), both velocity interval samples exhibit comparable $N$(H {\sc i}) distributions and show a clear decreasing trend of $N$(H {\sc i}) with increasing $f_{\ce{H2}}^N$. This trend is particularly pronounced in the 0.1 $f_{\ce{H2}}^N$-binned averages (red and blue markers). The decline in $N$(H {\sc i}) at $f_{\ce{H2}}^N \sim 0.5-0.6$, as revealed by the binned averages, provides strong evidence for the conversion of atomic hydrogen into \ce{H2} and other molecular species, potentially marking the transition from translucent cloud regimes to diffuse molecular gas. The derived $N$(H {\sc i}) range corresponds to a total hydrogen column density of $N$(H$_{\rm total}$) $\sim 3.15\times10^{20}-9.31\times10^{22}$ cm$^{-2}$. The simulation results presented in \citet{Bellomi2020_HtoH2} and \citet{Valdivia2016_H2_multiphase_MCs} show that similar ranges of $N$(H$_{\rm total}$) and $f_{\rm H_2}^N$  correspond to both warm and cold gas within intermixed two-phase diffuse ISM and predominantly trace cold and dense regions of diffuse molecular gas.

We also examined abundance variations of OH, \ce{HCO+}, CCH, and O with molecular gas fraction. Figure\,\ref{fig:x_fmol} shows the abundance scatter plots of the species relative to $N$(H$_{\rm total}$) (upper panel) and $N$(\ce{H2}) (lower panel) as a function of the molecular gas fraction. CH is employed for deriving \ce{H2} column densities, and consequently, it is excluded from the abundance comparisons presented in the plot. In the upper plot, the abundances of O exhibit relatively consistent values with the mean value $\langle X{\rm (O)} \rangle$ of $(3.09\pm0.64)\times10^{-4}$ across all phases, from atomic to molecular, via the transition. Conversely, oxygen abundances relative to neutral atomic hydrogen exhibit a notable increase with significant variability as the molecular gas fraction rises at the beginng of the molecular gas phase region (refer to Fig.\,\ref{appendix:abundances}), whereas the abundance of oxygen relative to molecular hydrogen shows a clear decline (as indicated in the lower plot of Fig.\,\ref{fig:x_fmol}). The black solid lines in Figs.\,\ref{fig:x_fmol} and \ref{appendix:abundances} represent the predicted oxygen abundances derived from $N$(O) values, assuming a constant oxygen abundance relative to $N$(H$_{\rm total}$) in both atomic and molecular gas. While these assumed $X$(O) generally agree with observational measurements, considerable scatter is observed in both comparisons in the transition phase area. This may suggest that the abundance of oxygen is not uniformly consistent in atomic and molecular gas regions along the lines of sight.

\begin{figure}[t]
    \centering
    \includegraphics[width=0.38\textwidth]{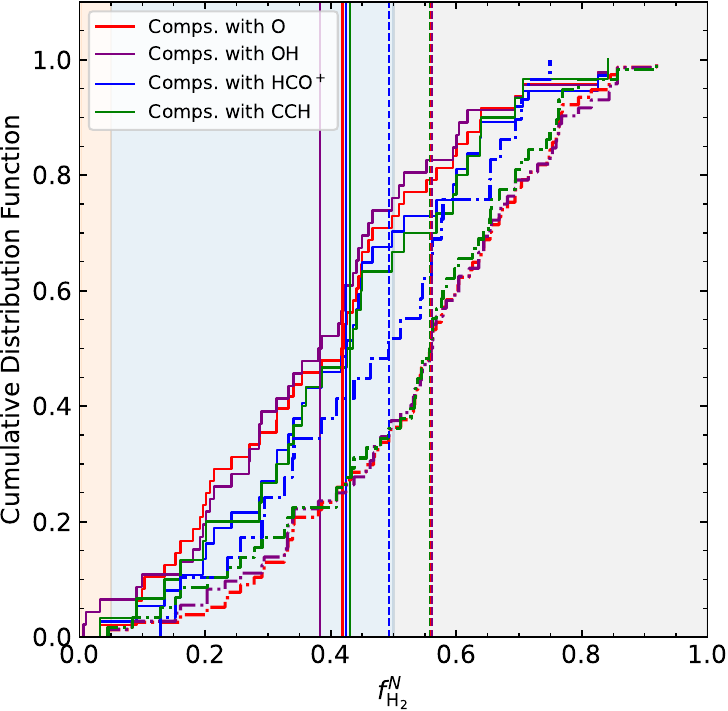}
    \caption{Cumulative distribution functions of absorption components exhibiting the presence of O (red), OH (magenta), \ce{HCO+} (blue), and CCH (bright green) as a function of molecular gas fraction. Solid lines correspond to velocity components defined by VI(A) intervals, while dash-dotted lines represent components defined by VI(M) intervals.  Vertical solid lines show the median values of $f_{\ce{H2}}^N$ for the four species in the VI(A) sample, while dashed lines show those for the VI(M) sample. Shaded colored regions denote distinct gas phases as indicated in Fig.\,\ref{fig:x_fmol}. } 
    \label{fig:fmol_cdf}
\end{figure}
In the upper plot of Fig.~\ref{fig:x_fmol}, the abundances of OH, \ce{HCO+}, and CCH increase markedly, by approximately two orders of magnitude, as the molecular gas fraction rises. The color-coded solid lines represent the predicted abundances of these species, calculated under the assumption of constant abundances relative to \ce{H2}. Specifically, the column densities $N$(X) were derived using the median ratios $N$(OH)/$N$(\ce{H2})~$=1.2\times10^{-7}$, $N$(\ce{HCO+})/$N$(\ce{H2})~$=3.0\times10^{-9}$, and $N$(CCH)/$N$(\ce{H2})~=$3.1\times10^{-8}$, and the abundances determined here are in great agreement with the previous observations ($N$(OH)/$N$(\ce{OH})~$=1.0\times10^{-7}$ and $1.2\times10^{-7}$; \citealt{Nguyen2018_OH_dust, Rugel2025_OH_HI}, $N$(CCH)/$N$(\ce{H2})~$= (3.2\pm1.1)\times10^{-8}$ and $(2.9\pm1.3)\times10^{-8}$; \citealt{Gerin2010_ch_cch,Lucas2000_cch_c3h2}, and $N$(\ce{HCO+})/$N$(\ce{H2})~$=3-6\times10^{-9}$; \citealt{Lucas1996_HCOp}). These ratios, obtained from our observational data, are indicated as horizontal solid lines in the lower panel of Fig.~\ref{fig:x_fmol}. The close agreement between the predicted abundances relative to H$_{\rm total}$, \ce{H2} (solid curves in Fig.~\ref{fig:x_fmol}), and H~{\sc i} (solid curves in  Fig.~\ref{appendix:abundances}) and the observational data for OH, \ce{HCO+}, and CCH strongly indicates a tight correlation between molecular hydrogen and the formation of these species. In addition, the abundance curves for CCH and \ce{HCO+} decline noticeably at $f_{\ce{H2}}^N \leq 0.1$–0.15, whereas the OH abundance drops more sharply at lower molecular fractions, around $f_{\ce{H2}}^N < 0.05$. Within the outskirts of the Chamaeleon cloud, \ce{HCO+} absorption was not detected toward positions where $f_{\ce{H2}}^N \leq 0.06$ \citep{Liszt2023_DNM}, roughly consistent with the abundance drop observed in the upper plot of Fig.~\ref{fig:x_fmol}. Whereas the more pronounced decline of the predicted OH abundance curve at lower molecular gas fractions, some OH data points, compared with CCH and \ce{HCO+}, are still higher, suggesting that OH absorption can occur toward relatively more diffuse regions with lower molecular gas fractions. This behavior further supports the interpretation of OH as an effective tracer of CO-dark molecular gas. The median abundance of CCH relative to \ce{H2} ($3.1\times10^{-8}$) is remarkably close to that of CH ($N$(CH)/$N$(\ce{H2})~$= 3.5\times10^{-8}$). Despite the large scatter observed in CCH abundances across the range of $f_{\ce{H2}}^N$ compared with those of OH and \ce{HCO+}, the consistency between our measurements and earlier studies supports the close chemical relationship between CH and CCH in the diffuse and translucent ISM. In addition, as shown in the lower panel of Fig. \ref{fig:x_fmol}, the abundances of OH and \ce{HCO+} remain relatively constant across a range of molecular gas fractions, while the abundance of atomic oxygen decreases. This observation suggests that it is difficult to account for the consumption of oxygen in the formation of OH and \ce{HCO+} as the molecular gas fraction (\(f_{\ce{H2}}^N\)) increases. However, we cannot rule out the possibility that oxygen may be converting into \ce{H2O} and CO.

In the column density comparisons of OH, CH, \ce{HCO+}, and CCH, no significant differences are observed between the two velocity-interval samples, nor in the relation between oxygen column density and $N(\ce{H2})$. However, the $N(\mathrm{O})$ values associated with the VI(A) sample are systematically higher than those in the VI(M) sample. This suggests that the abundances of these species are influenced by the physical conditions distinguishing the two velocity-selected intervals. To investigate this further, we examined whether the VI(M) intervals preferentially trace gas with higher molecular fractions compared to the VI(A) intervals.

Figure~\ref{fig:fmol_cdf} shows the cumulative distribution functions of $f_{\ce{H2}}^N$ for the VI(A) (solid curves) and VI(M) (dot-dashed curves) samples for O, OH, \ce{HCO+}, and CCH. Kolmogorov–Smirnov (KS) tests were performed to assess whether the distributions differ significantly. The small $p$-values (ranging from $9.6\times10^{-6}$ to $5.5\times10^{-5}$) indicate that the null hypothesis can be rejected for O and OH. This implies that the O and OH absorption components within the \ce{HCO+} velocity intervals are clearly associated with regions of higher molecular gas fraction. In other words, the VI(A) intervals encompass a larger volume of low-density atomic gas, leading to smaller molecular gas fractions when integrated over the total column density compared to those obtained from the VI(M) intervals. For \ce{HCO+} and CCH, the KS test results yield higher $p$-values of 0.46 and 0.023, respectively. Therefore, we cannot reject the null hypothesis that the \ce{HCO+} and CCH samples integrated over the VI(A) and VI(M) velocity intervals originate from the same underlying distributions. The slight differences observed in their cumulative distributions likely result from variations in $N$(\ce{H2}), which is primarily influenced by CH and, to a lesser extent, by OH and $N$(H~{\sc i}). 

The median $f_{\ce{H2}}^N$ values for the VI(A) intervals are around 0.4 (solid vertical lines in Fig.~\ref{fig:fmol_cdf}), whereas those for the VI(M) intervals are approximately 0.5 or higher (dashed vertical lines). Numerical simulations incorporating shocks \citep[e.g.,][]{Komichi2024_simulation_3mmlines} and chemical models considering photodissociation regions (PDRs) \citep[e.g.,][]{Rybarczyk2022_HCOp_COdark,Kim2023_hygal2} indicate that \ce{HCO+} predominantly arises in diffuse molecular and translucent cloud gas ($n \sim$ a few hundreds to thousands cm$^{-3}$). Comparisons of \ce{HCO+} absorption and emission lines in \citet{Liszt2025_HCOp} also suggest that these tracers probe gas with $n \approx 50 - 200$ cm$^{-3}$, denser than the unstable, non-equilibrium gas ($n \sim 1-8$ cm$^{-3}$) where significant \ce{H2} formation can still occur \citep[e.g.,][]{Wolfire2003_UNM,Bialy2019}. This implies that using the \ce{HCO+} absorption velocity intervals defined here does not perfectly trace full CO-dark \ce{H2} gas ($n \lesssim 30$ cm$^{-3}$), though it performs slightly better than CO-based tracers \citep[e.g.,][]{Liszt2019_co-dark_alma_hcop,Rybarczyk2022_HCOp_COdark}. Furthermore, this clearly indicates that it is necessary to compare these species to fully understand their relationships and the physical conditions, including the pathways of their formation and destruction, across the various diffuse ISM environments.

\section{Summary and conclusions}
\label{sec:summary}
Using SOFIA observations, complemented by ancillary data from the IRAM 30 m telescope and ALMA-ACA, we investigated correlations among the known \ce{H2} proxies CH, OH, \ce{HCO+}, and CCH, and abundance variations of [O~{\sc i}] in absorption toward 25 submillimeter-bright Galactic star-forming regions, most of which are Galactic H{\sc ii} regions. All four species (CH, OH, \ce{HCO+}, and CCH) exhibit strong mutual correlations, with Spearman correlation coefficients ranging from 0.72 to 0.87. The only exception is the $N$(OH)–$N$(CCH) relation, for which the limited number of data points at low column densities prevents robust correlation analysis. The derived average column density ratios among these species are $N$(OH)/$N$(CH) = $3.29 \pm 0.11$, $N$(\ce{HCO+})/$N$(CH) = $0.085 \pm 0.006$, $N$(CCH)/$N$(CH) = $1.02 \pm 0.07$, $N$(\ce{HCO+})/$N$(OH) = $0.02\pm0.01$, and $N$(CCH)/$N$(OH) = $0.25 \pm 0.01$. These ratios are in good agreement with previous measurements in local diffuse clouds.

Notably, the OH column density shows particularly tight correlations with \ce{HCO+} and CCH compared with CH. Similar trends were also reported by \citepalias{Kim2023_hygal2}, suggesting that these relationships may reflect the underlying chemical and physical connections among the species in their formation and destruction pathways. The column density ratios among these \ce{H2} proxies appear remarkably consistent across different Galactic environments and remain uniform and independent of the velocity intervals considered (i.e., VI(A) and VI(M)). The oxygen abundance relative to total hydrogen is $(3.09\pm0.64)\times10^{-4}$, which is broadly consistent with previous observational estimates \citep[e.g.,][]{Cartledge2004,Lis2023}, though with noticeable scatter. The oxygen abundances derived from the VI(A) intervals generally fall below the elemental solar value of $4.9\times10^{-4}$. 

Using CH as a tracer to convert to \ce{H2} column densities, together with H~{\sc i} absorption-line data, we investigated the molecular gas fractions along the HyGAL lines of sight. The $N$(H~{\sc i}) values decrease toward regions where $f_{\ce{H2}}^N$ exceeds 0.5, marking the transition to the molecular gas phase. The oxygen abundances relative to total hydrogen remain roughly constant across both low and high molecular fraction regimes. In contrast, the abundances of OH, \ce{HCO+}, and CCH increase distinctly with $f_{\ce{H2}}^N$. In regions with low $f_{\ce{H2}}^N$, the abundance of OH remains relatively high compared to other species, such as \ce{HCO+} and CCH. A comparison between the VI(A) and VI(M) samples reveals that the latter, integrated over the \ce{HCO+} absorption-line intervals, traces gas regions with higher molecular fractions ($f_{\ce{H2}}^N > 0.5$), whereas the VI(A) sample peaks around $f_{\ce{H2}}^N \sim 0.4$. These results indicate that \ce{HCO+} absorption lines can trace regions of diffuse molecular gas with high molecular gas fractions, while OH absorption lines can trace areas with lower molecular gas fractions. 

\begin{acknowledgements}
 This work is based on observations made with the NASA/DLR Stratospheric Observatory for Infrared Astronomy (SOFIA). SOFIA is jointly operated by the Universities Space Research Association, Inc. (USRA), under NASA contract NNA17BF53C, and the Deutsches SOFIA Institut (DSI) under DLR contract 50 OK 0901 to the University of Stuttgart. W.-J. K. was supported by DLR/Verbundforschung Astronomie und Astrophysik Grant 50 OR 2007 for this work. We gratefully acknowledge the excellent support provided by the SOFIA Operations Team and the GREAT Instrument Team. Part of this research was carried out at the Jet Propulsion Laboratory, California Institute of Technology, under a contract with the National Aeronautics and Space Administration (80NM0018D0004). D.C.L. acknowledges financial support from the National Aeronautics and Space Administration (NASA) Astrophysics Data Analysis Program (ADAP). A.S-M.\ acknowledges support from the PID2023-146675NB grant funded by MCIN/AEI/10.13039/501100011033, and by the programme Unidad de Excelencia Mar\'{\i}a de Maeztu CEX2020-001058-M, as well as support from the RyC2021-032892-I grant funded by MCIN/AEI/10.13039/501100011033 and by the European Union `Next GenerationEU'/PRTR. D. S. acknowledges support of the Bonn-Cologne Graduate School, which is funded through the German Excellence Initiative as well as funding by the Deutsche Forschungsgemeinschaft (DFG) via the Collaborative Research Center (CRC) SFB 1601 ``Habitats of massive stars across cosmic time'' (subprojects B1 and B4 ). F. W. acknowledges support by the DFG via the CRC SFB 1601, subprojects B1. W.-J. K. acknowledges support by the DFG via the CRC SFB 1601, subprojects A2 and B1. V. O.-O. acknowledges support of the DFG via the CRC SFB 1601 subproject A6. P. S. acknowledges support by the DFG via the CRC SFB 1601, subprojects A1 and A2. S.B. acknowledges support from the ISF grant number 2071540, the GIF grant number I-1568-303.7/2024, the NSF-BSF grant number 2023761, and the Alon Fellowship prize for junior faculty. 
\end{acknowledgements}

\bibliographystyle{aa}
\bibliography{aa58863-26}

\newpage
\onecolumn

\begin{appendix}
\section{Source information and ALMA archival data}
Table \ref{tab:hygal_complete_source_list} summarizes the source coordinates, systemic velocities, kinematic distances, and Galactocentric distances of all HyGAL sources.

\begin{table*}[h!]
\begin{center}
\small
\caption{The summary of HyGAL source information. }
\label{tab:hygal_complete_source_list}
\begin{tabular}{ll cc rr r c r l}
\hline\hline
\# &Source & Right Ascension & Declination & Gal. Long. & Gal. Lat.  & $\upsilon_{\rm LSR}$ & \multicolumn{1}{c}{$d$} & $R_{\rm GAL}$  \\
&Designation & [hh:mm:ss] & [dd:mm:ss] & \multicolumn{1}{c}{[deg]} & \multicolumn{1}{c}{[deg]} & [km~s$^{-1}$]& [kpc] & [kpc] \\
\hline 
a & HGAL284.015$-$00.86 &	10:20:16.1	& $-$58:03:55.0 & 284.016  & $-$0.857  & 9.0  & 5.7& 9.0\\ 
b & HGAL285.26$-$00.05  &	10:31:29.5	& $-$58:02:19.5 & 285.263  & $-$0.051  & 3.4  & 4.3 & 8.2\\ 
c & G291.579$-$00.431   &	11:15:05.7	& $-$61:09:40.8 & 291.579  & $-$0.431  & 13.6 & 8.0 & 9.3\\ 
d & IRAS~12326$-$6245   &	12:35:35.9	& $-$63:02:29.0 & 301.138  & $-$0.225  & $-$39.3 & 4.6 & 7.2\\ 
e &G327.3$-$00.60 	    &	15:53:05.0	& $-$54:35:24.0 & 327.304  & $-$0.551  & $-$46.9 & 3.1 & 6.2\\ 
f & G328.307$+$0.423    &	15:54:07.2	& $-$53:11:40.0 & 328.309  & $+$0.429    & $-$93.6 & 5.8 & 4.6\\ 
g & IRAS~16060$-$5146   &	16:09:52.4	& $-$51:54:58.5 & 330.953  & $-$0.182  & $-$91.2 & 5.3 & 4.5\\ 
h & IRAS~16164$-$5046   &	16:20:11.9	& $-$50:53:17.0 & 332.827  & $-$0.551  & $-$57.3 & 3.6 & 5.4\\ 
i & IRAS~16352$-$4721   &	16:38:50.6	& $-$47:28:04.0 & 337.404  & $-$0.403  & $-$41.4  & 12.3 & 5.1\\ 
j & IRAS~16547$-$4247   &	16:58:17.2	& $-$42:52:08.9 & 343.126  & $-$0.063  & $-$30.6 & 2.7 & 5.8\\ 
k & NGC~6334~I  	    &	17:20:53.4	& $-$35:47:01.5 & 351.417  & $+$0.645    & $-$7.4  & 1.3 & 7.0\\ 
l & G357.558$-$00.321   &	17:40:57.2	& $-$31:10:59.3 & 357.557  & $-$0.321  & 5.3  & 9.0--11.8 & 1.0--3.6\\
m & HGAL0.55$-$0.85     &	17:50:14.5	& $-$28:54:30.7 & 0.546    & $-$0.851  & 16.7 & 7.7--9.2 & 0.4--1.0 \\
n & G09.62$+$0.19       &	18:06:14.9	& $-$20:31:37.0 & 9.620    & $+$0.194    & 4.3  & 5.2 & 3.3 \\
o & G10.47$+$0.03       &	18:08:38.4	& $-$19:51:52.0 & 10.472   & $+$0.026    & 67.6    & 8.6 & 1.6 & \\
p & G19.61$-$0.23       &	18:27:38.0	& $-$11:56:39.5 & 19.608   & $-$0.234  & 40.8    & 12.6 & 4.7 \\
q & G29.96$-$0.02       &	18:46:03.7	& $-$02:39:21.2 & 29.954   & $-$0.016  & 97.2    & 6.7 & 4.5 \\
r & G31.41+0.31         &	18:47:34.1	& $-$01:12:49.0 & 31.411   & $+$0.307    & 98.2    & 4.9 & 5.0 \\
s & W43~MM1             &	18:47:47.0	& $-$01:54:28.0 & 30.817   & $-$0.057  & 97.8    & 5.5 &5.0 \\
t & G32.80+0.19         &	18:50:30.6	& $-$00:02:00.0 & 32.796   & $+$0.191    & 14.6    & 13.0 & 7.4 \\ 
u & G45.07+0.13         &	19:13:22.0	& +10:50:54.0   & 45.071   & $+$0.133    & 59.2    & 4.3 & 6.2 \\
v & DR21                &   20:39:01.6  & +42:19:37.9   & 81.681   & $+$0.537     & $-$4.0  & 1.5 & 7.4 \\
w &  NGC~7538~IRS1      &   23:13:45.3  &  +61:28:11.7  & 111.542  & $+$0.777     & $-$59.0 & 2.6 &  9.8 \\
x &  W3~IRS5            &   02:25:40.5  &  +62:05:51.0  & 133.715  & $+$1.215     & $-$39.0 & 2.3 &  9.9\\
y &  W3(OH)             &   02:27:04.1  &  +61:52:22.1  & 133.948  & $+$1.064     & $-$48.0 & 2.0 &  9.6 \\
         \hline 
    \end{tabular}
    \end{center}
    \tablefoot{The information on this table is taken from Table\,2 of \citetalias{jacob2022_hygal}.}
\end{table*}
 
\begin{table*}[h!]
\small
\begin{center}
\caption{Summary of the ALMA-ACA archival data for \ce{HCO+} and CCH and detections. }
\label{tab:alma_archival}
    \begin{tabular}{c c c c c c c c}
    \hline \hline
     Source  &  \multicolumn{3}{c}{\ce{HCO+}} & & \multicolumn{3}{c}{\ce{CCH}} \\ \cline{2-4} \cline{6-8}
             &  D & $T_{\rm c}$ [K] & $T_{\rm rms}$ [K] &  & D & $T_{\rm c}$ [K] & $T_{\rm rms}$ [K] \\ 
    \hline
    IRAS~12326$-$6245 &  Y & 0.7 & 0.013 &  & Y & 0.6 & 0.014 \\ 
    G328.307$+$0.423  &  Y & 1.0 & 0.020 &  & Y & 9.5 & 0.019 \\ 
    IRAS~16060$-$5146 & Y & 1.8 & 0.025 &  & Y & 1.7 & 0.021 \\ 
    IRAS~16164$-$5046 & Y & 1.3 & 0.022 &  & Y & 1.2 & 0.019 \\ 
    IRAS~16352$-$4721 & Y & 0.1 & 0.029 &  & N & 0.1 & 0.023 \\ 
    IRAS~16547$-$4247 &  Y & 0.1 & 0.019 &  & N & 0.1 & 0.015 \\ 
    \hline
    \end{tabular}    
    \tablefoot{Column D exhibits the detections of absorption features against the observed background continuum sources. ``Y'' indicates a detection, while for the non-detections (N), tentative absorption features are present, but the confirmation of these features remains ambiguous due to poor signal-to-noise ratios. }
\end{center}
\end{table*}
Table \ref{tab:alma_archival} summarizes the ALMA-ACA archival data for \ce{HCO+} and CCH and their detections.
\newpage
\section{Spectral line plots}\label{Appendix:spectral}
Figures \ref{appendix:spectra1} and \ref{appendix:spectra2} show the spectra of each observed species in each source, overlaid with the corresponding XCLASS model spectra.

\begin{figure*}[h!]
    \centering
    \includegraphics[width=0.23\textwidth]{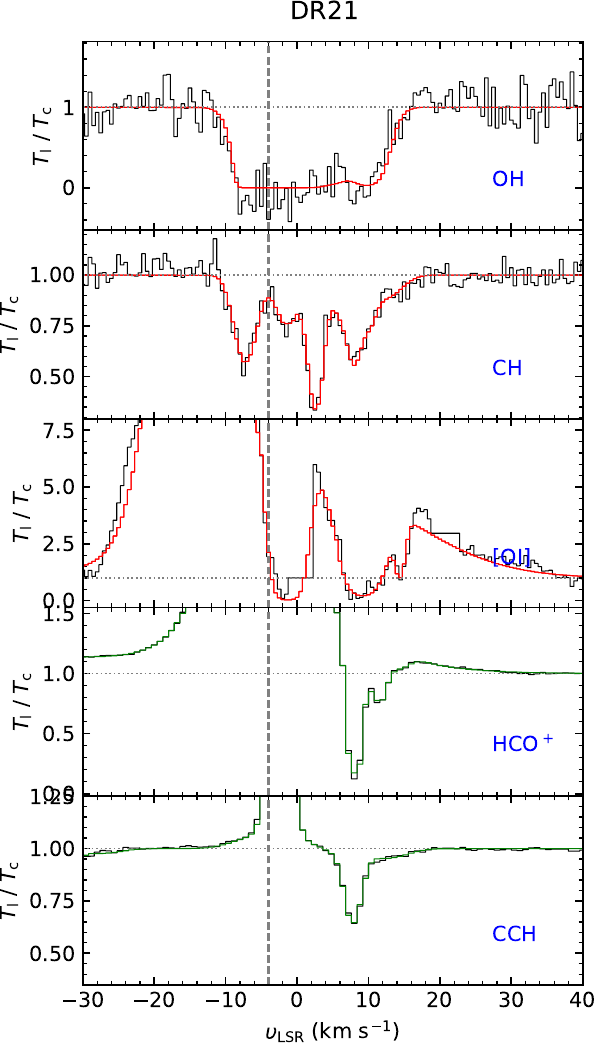}
    \includegraphics[width=0.22\textwidth]{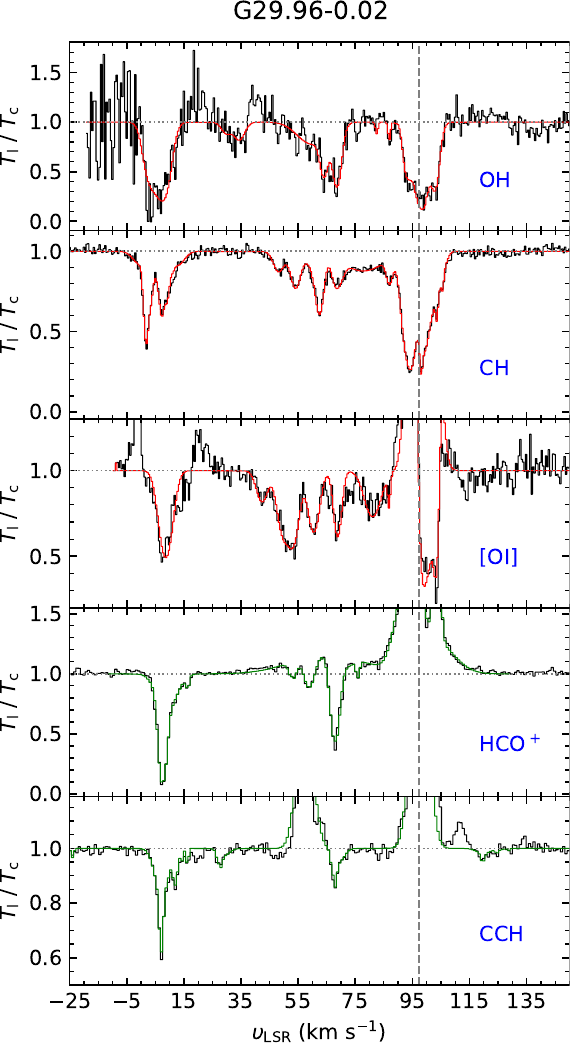}
    \includegraphics[width=0.22\textwidth]{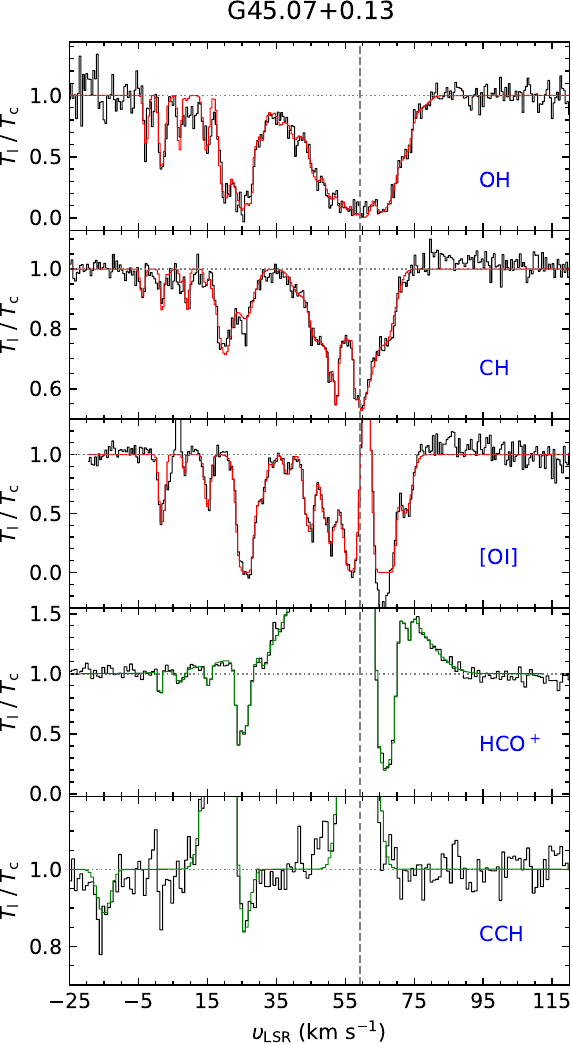}
    \includegraphics[width=0.22\textwidth]{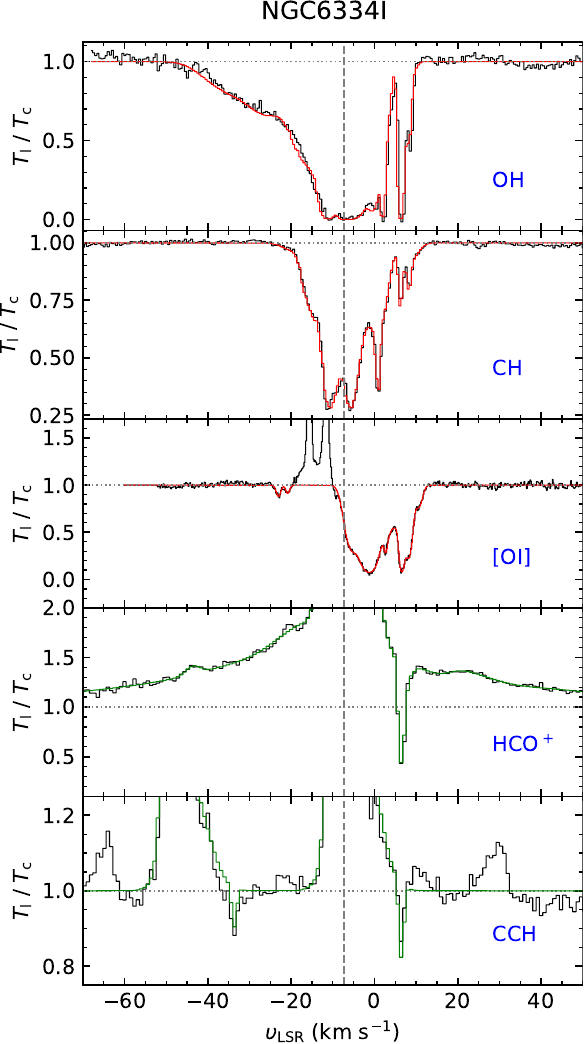}
    \includegraphics[width=0.23\textwidth]{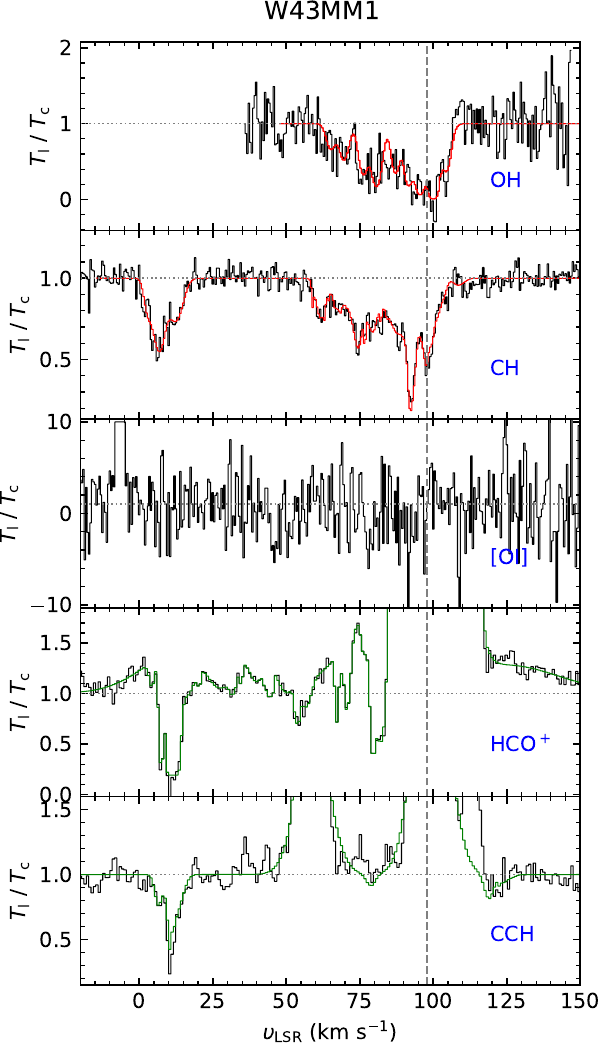}
    \includegraphics[width=0.22\textwidth]{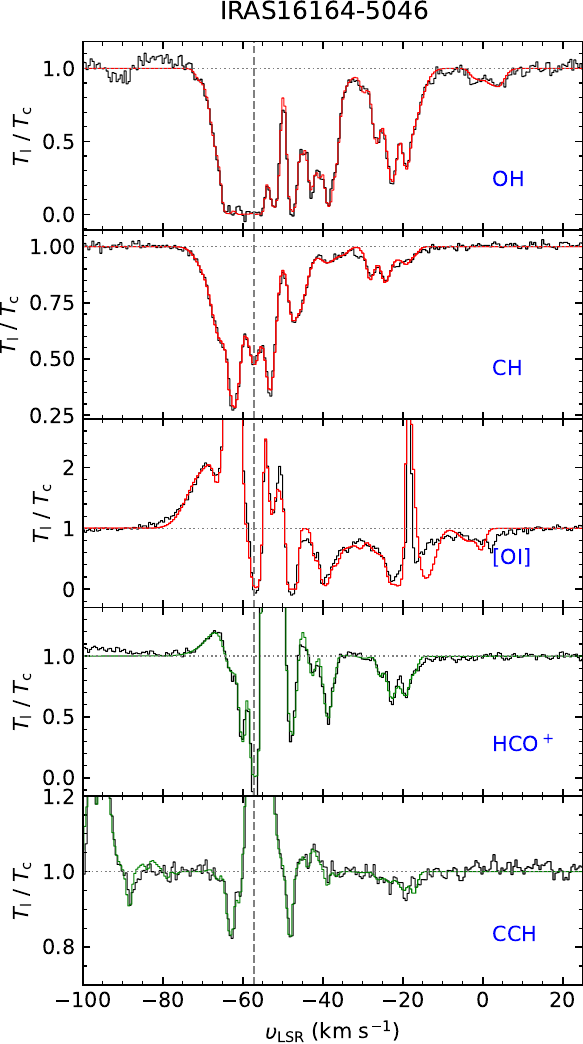}
    \includegraphics[width=0.22\textwidth]{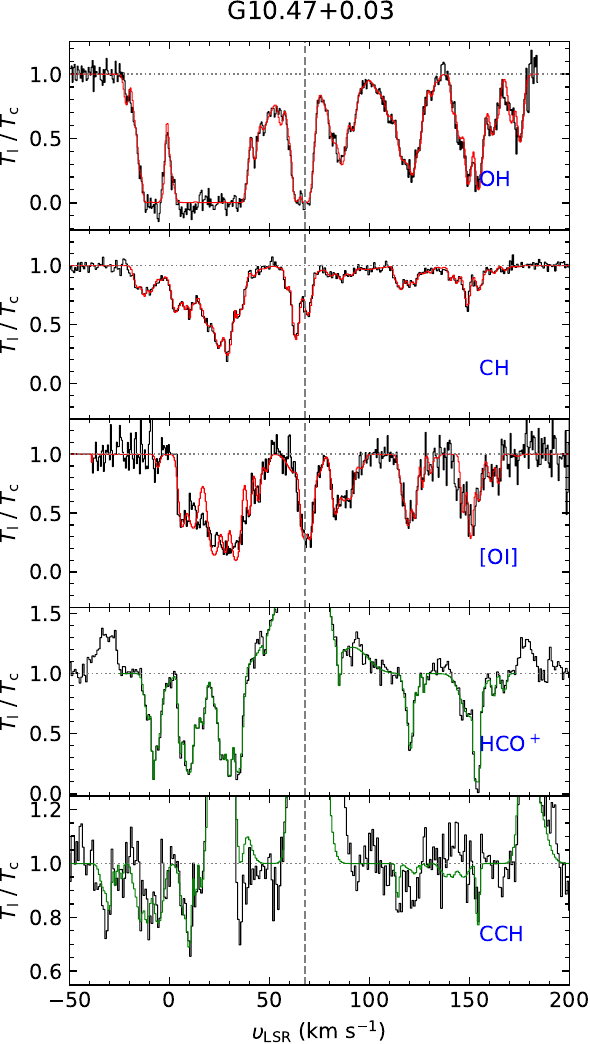}
    \includegraphics[width=0.22\textwidth]{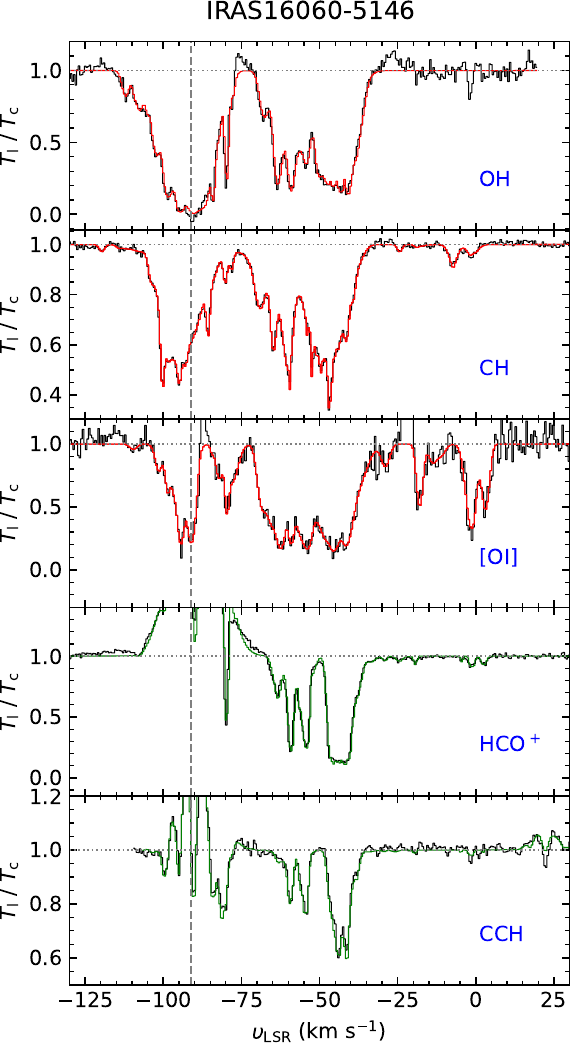}
   \includegraphics[width=0.22\textwidth]
    {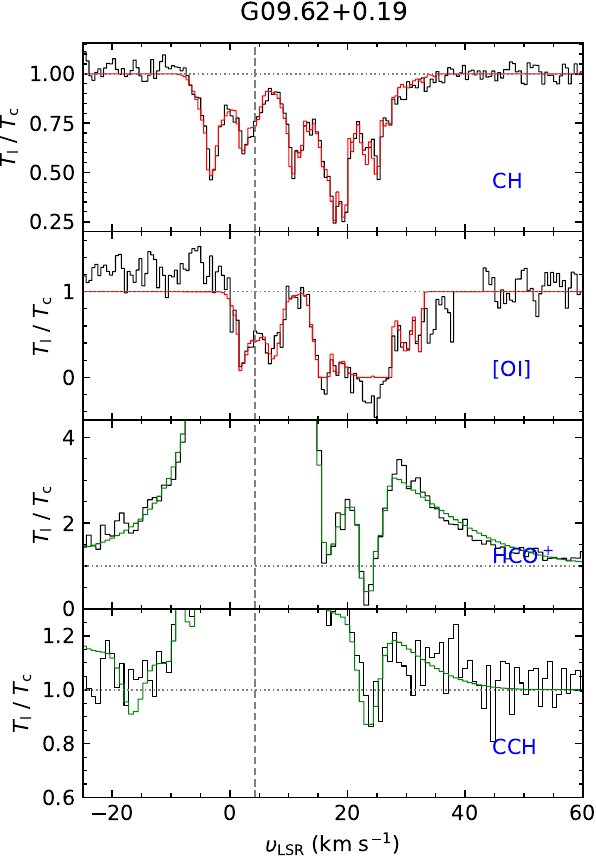}
    \includegraphics[width=0.22\textwidth]{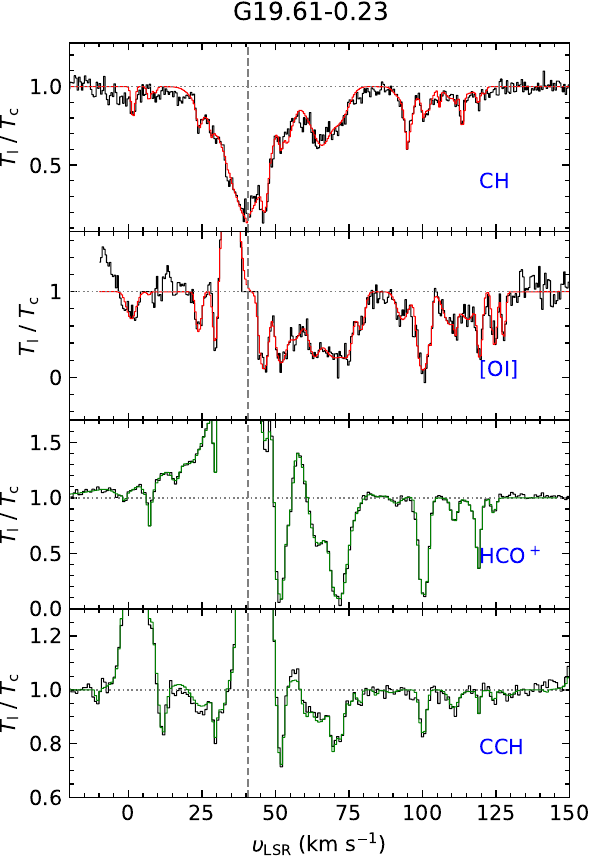}
    \includegraphics[width=0.22\textwidth]
    {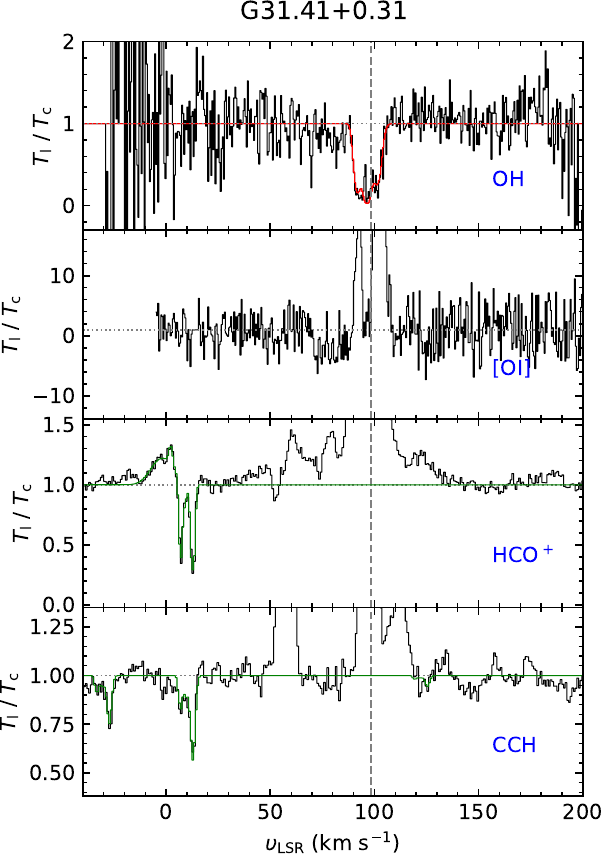}
    \includegraphics[width=0.22\textwidth]{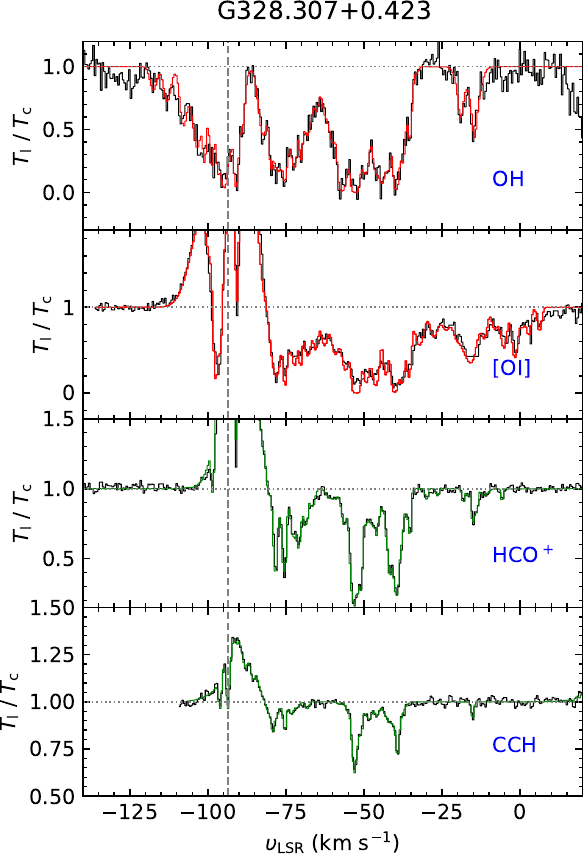}
    \caption{Same as Fig.\,\ref{fig:spectra1}. } 
    \label{appendix:spectra1}
\end{figure*}


\begin{figure*}
    \centering
    \includegraphics[width=0.22\textwidth]{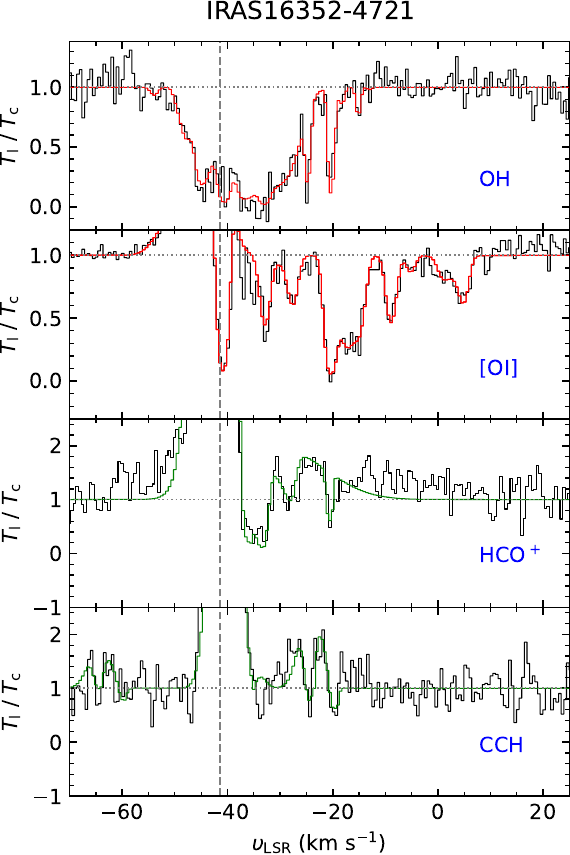}
    \includegraphics[width=0.22\textwidth]{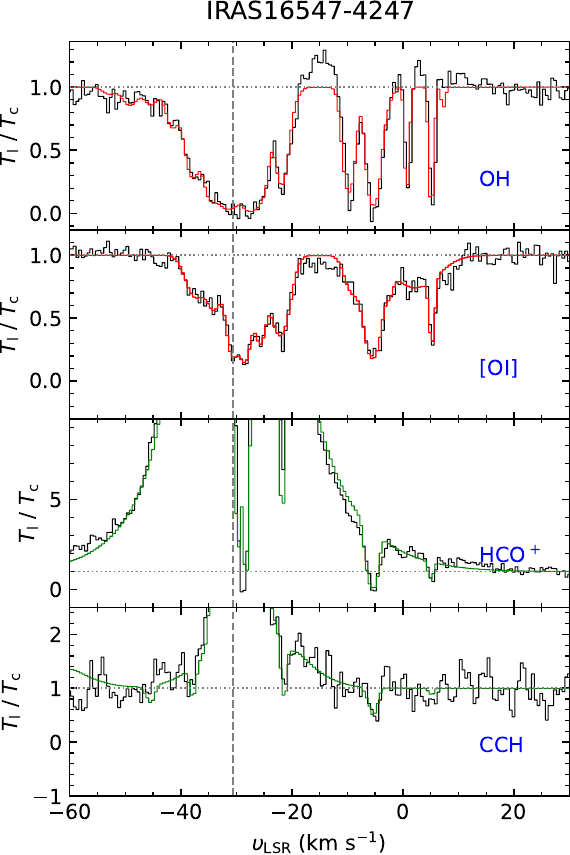}
    \includegraphics[width=0.23\textwidth]{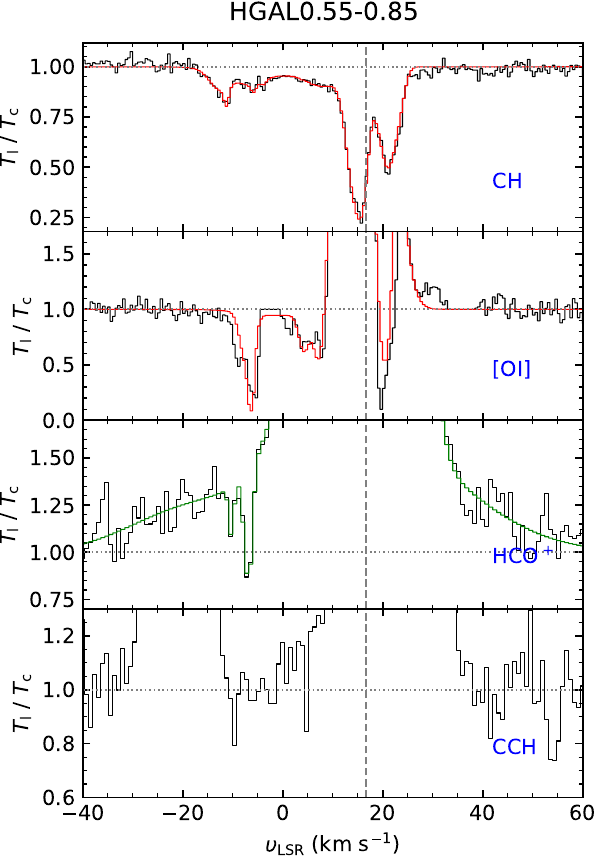}
    \includegraphics[width=0.22\textwidth]{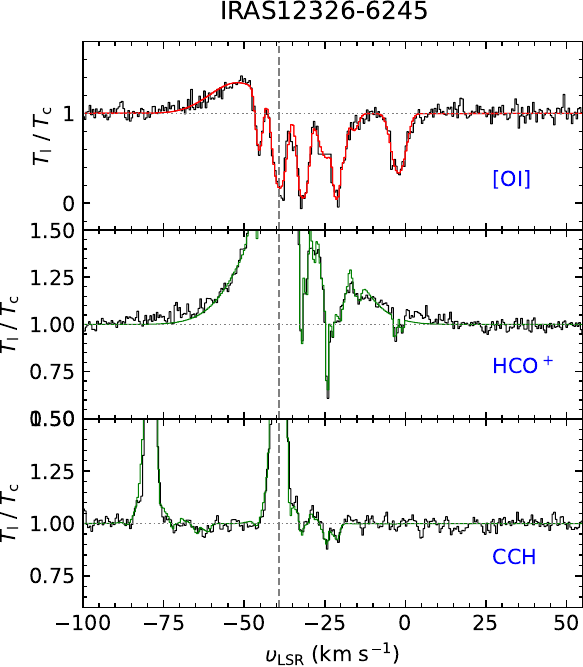}
    \includegraphics[width=0.22\textwidth]{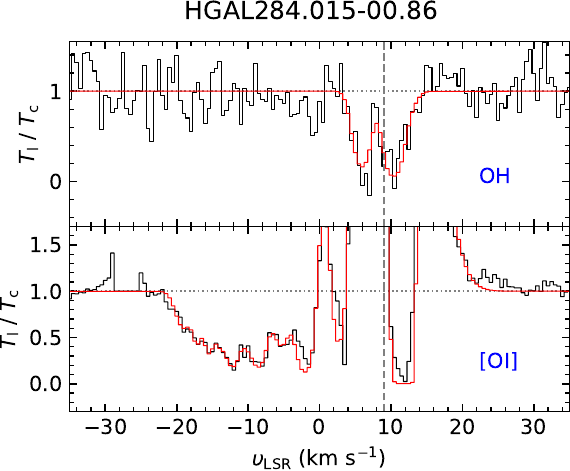}
    \includegraphics[width=0.22\textwidth]{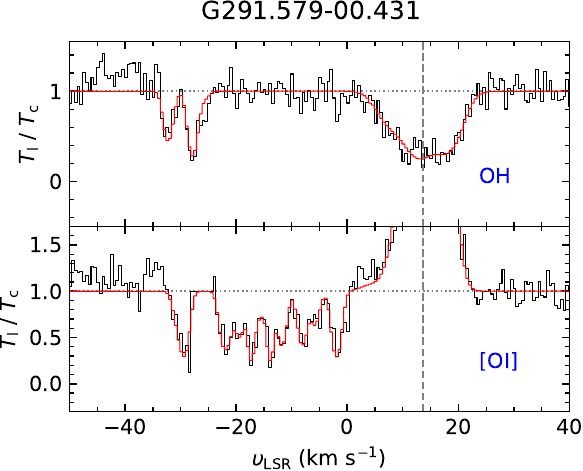}
    \includegraphics[width=0.22\textwidth]{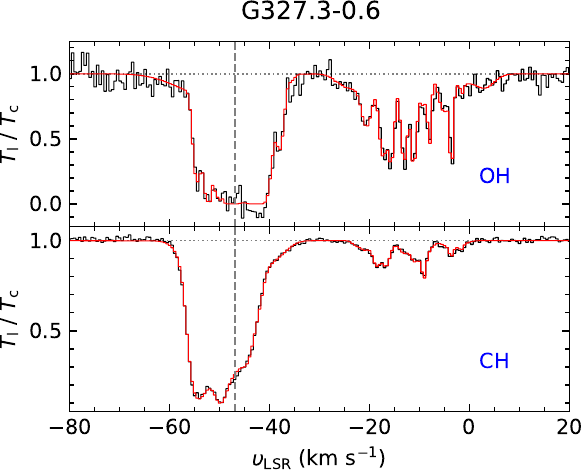}
    \includegraphics[width=0.22\textwidth]{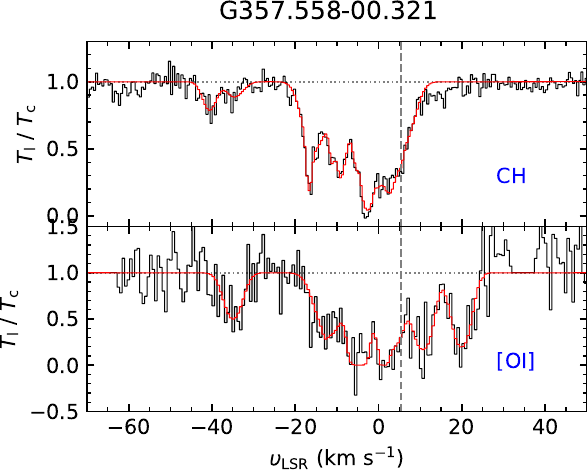}
    \includegraphics[width=0.22\textwidth]{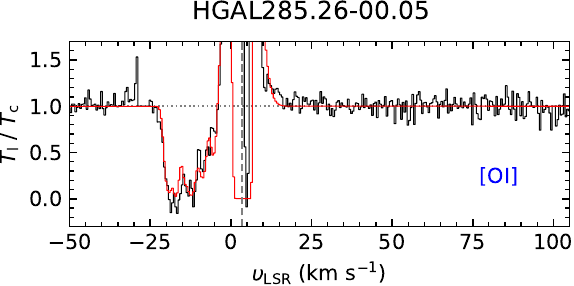}
    \caption{Same as Fig.\,\ref{fig:spectra1}.}
    \label{appendix:spectra2}
\end{figure*}

\section{Column density spectra}\label{Appendix:Ncol}
Figure\,\ref{Appendix:Ncol} shows the column density profile per velocity interval for each detected species in each observed source, and Table\ref{tab:column_density_tb1} lists the fraction of the full derived column densities over the specific velocity intervals for VI(A). Another column density table for the velocity interval for VI(M) is available at the CDS via anonymous ftp to cdsarc.cds.unistra.fr (130.79.128.5). 
\begin{figure*}[h!]
    \includegraphics[width=0.195\textwidth]{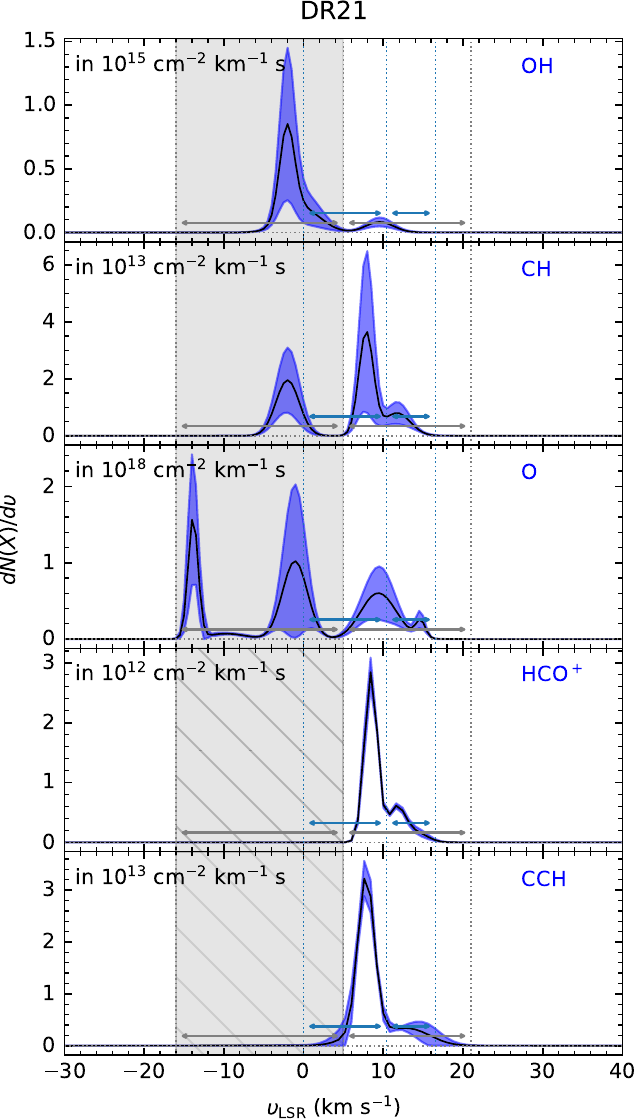}
    \includegraphics[width=0.195\textwidth]{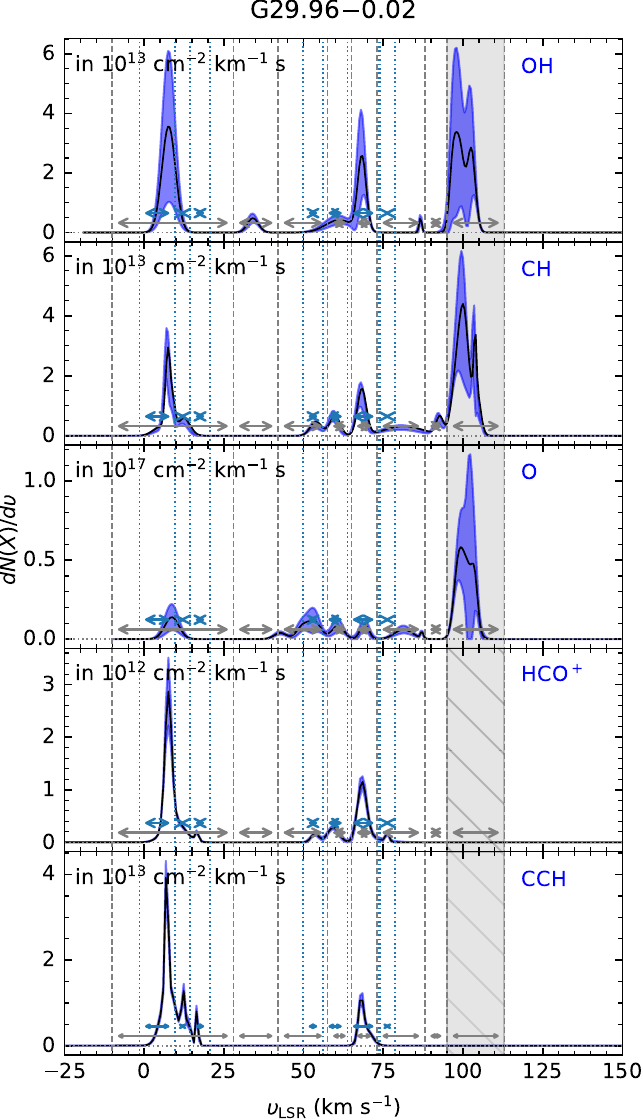}
    \includegraphics[width=0.195\textwidth]{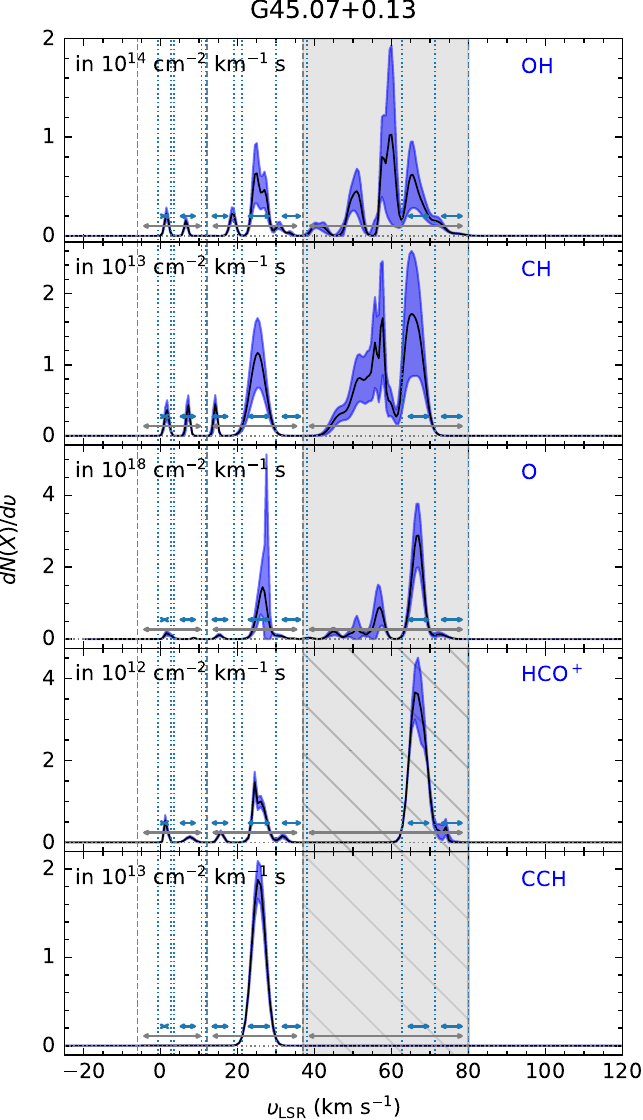}
    \includegraphics[width=0.190\textwidth]{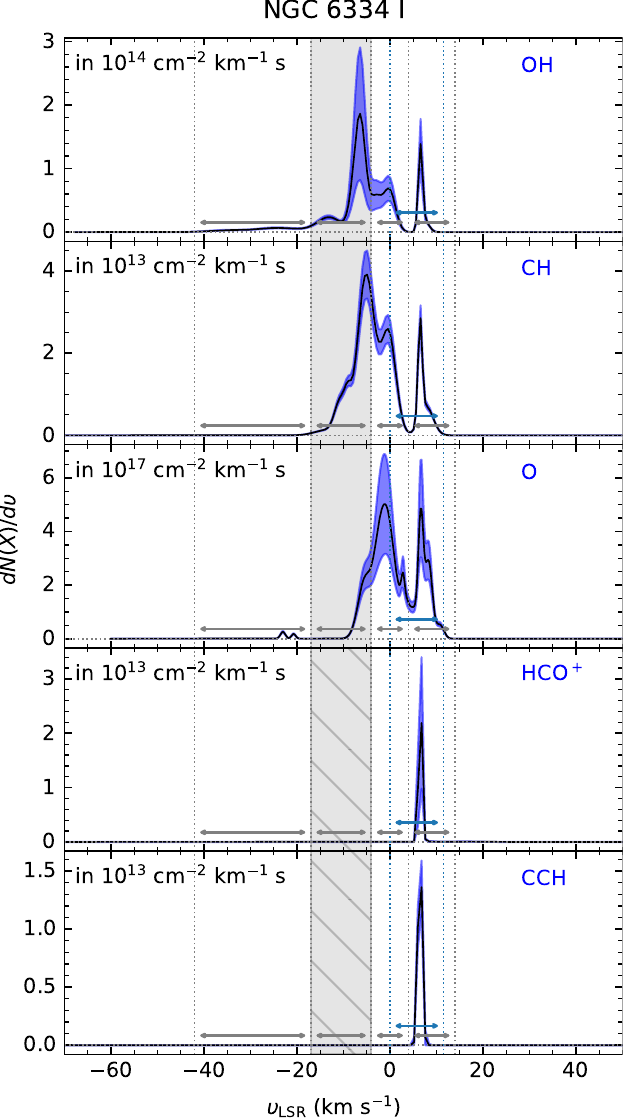}
    \includegraphics[width=0.195\textwidth]{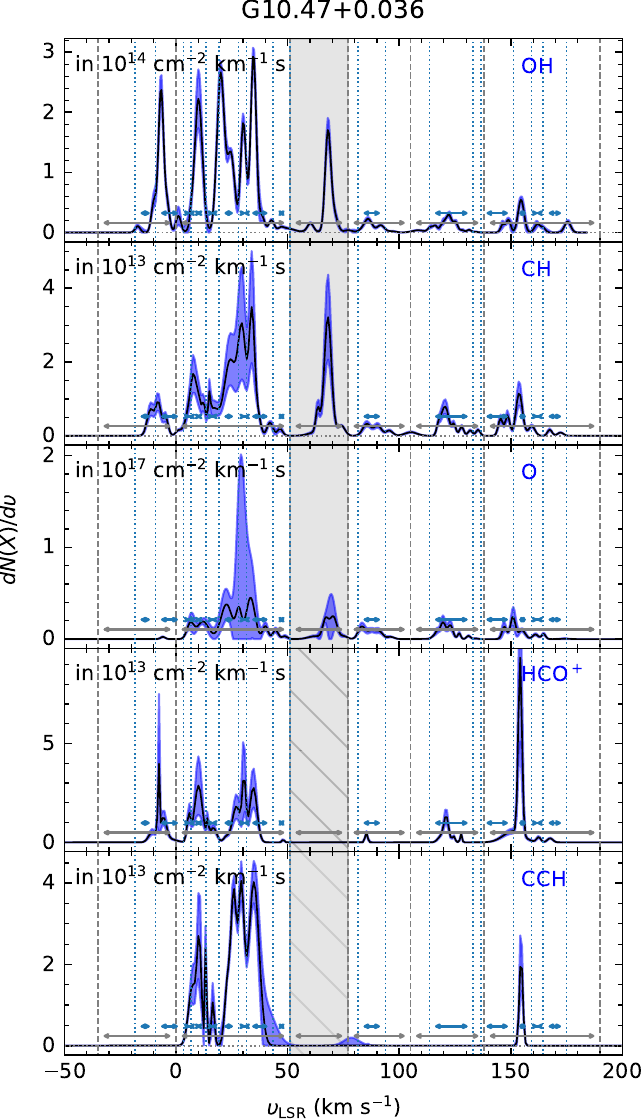}
    \includegraphics[width=0.195\textwidth]{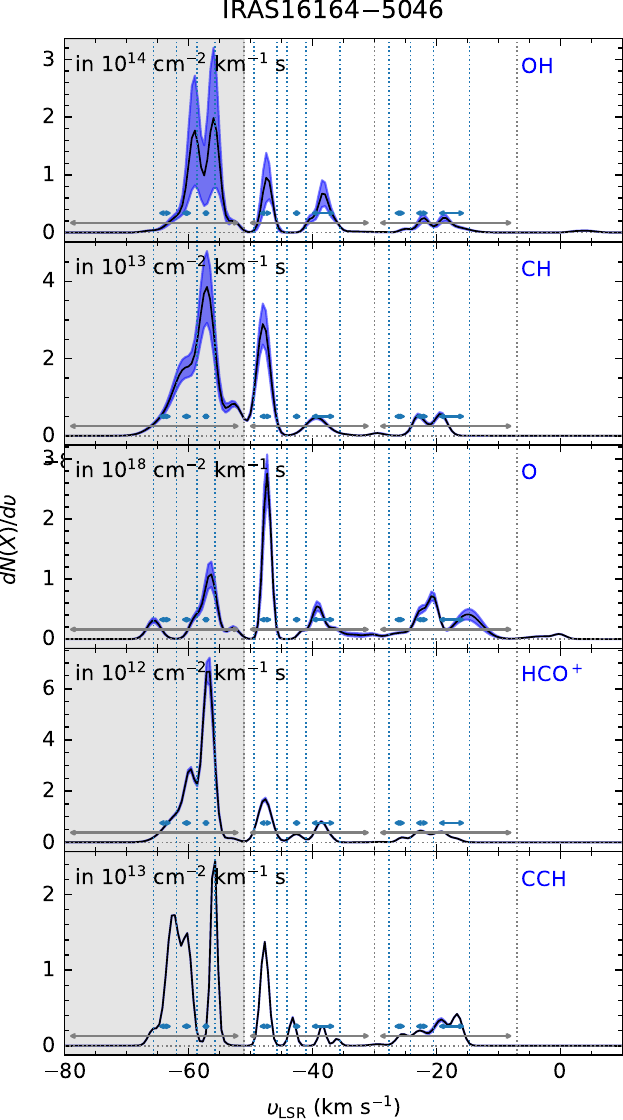}
    \includegraphics[width=0.195\textwidth]{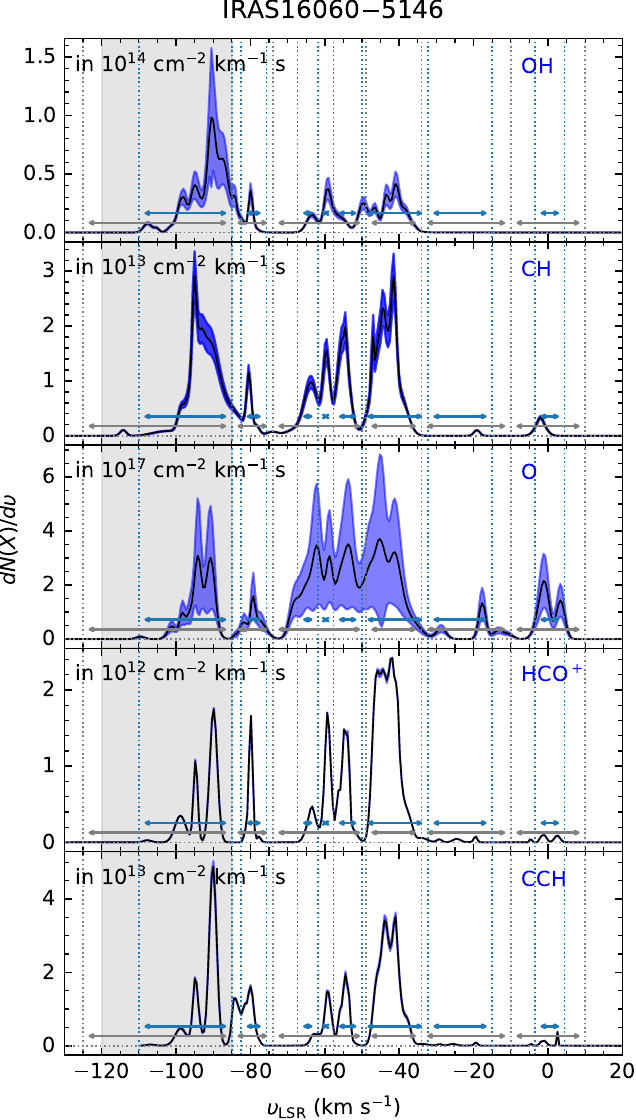}
    \includegraphics[width=0.195\textwidth]{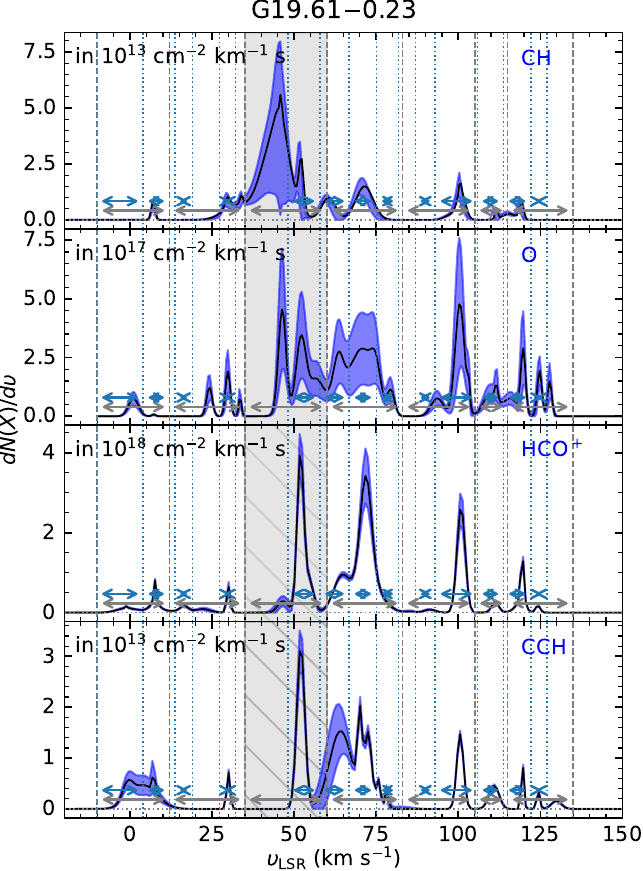}
    \includegraphics[width=0.195\textwidth]{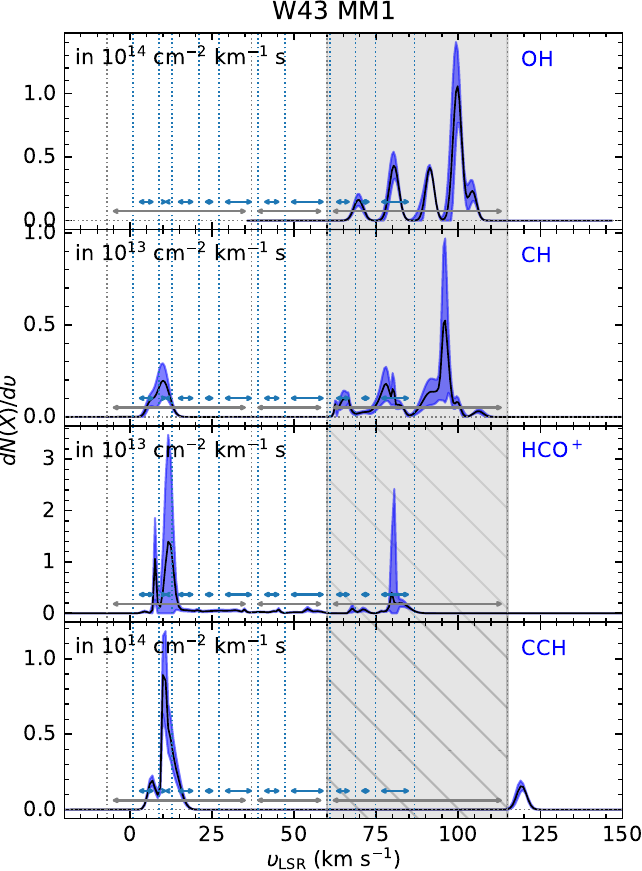}
    \includegraphics[width=0.195\textwidth]{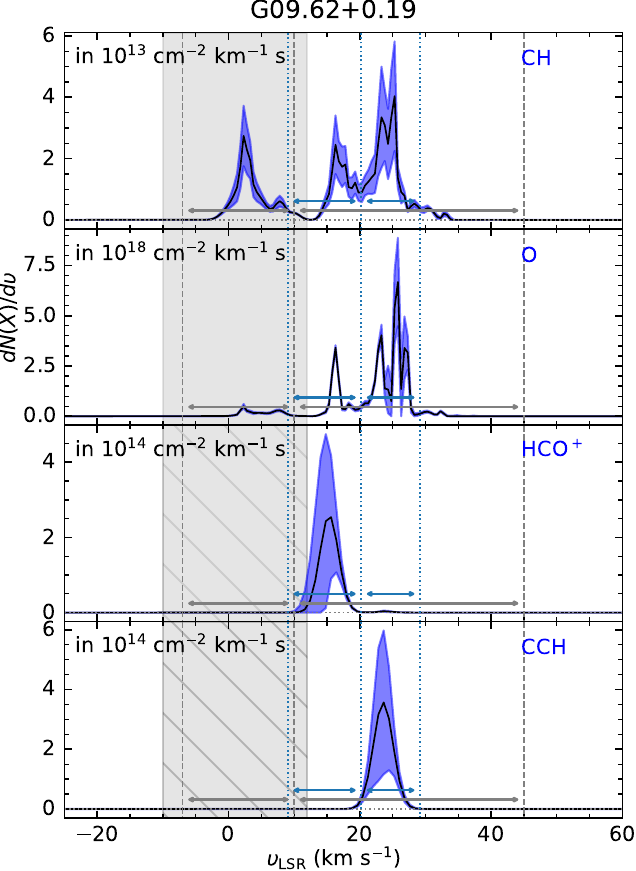}
    \includegraphics[width=0.195\textwidth]{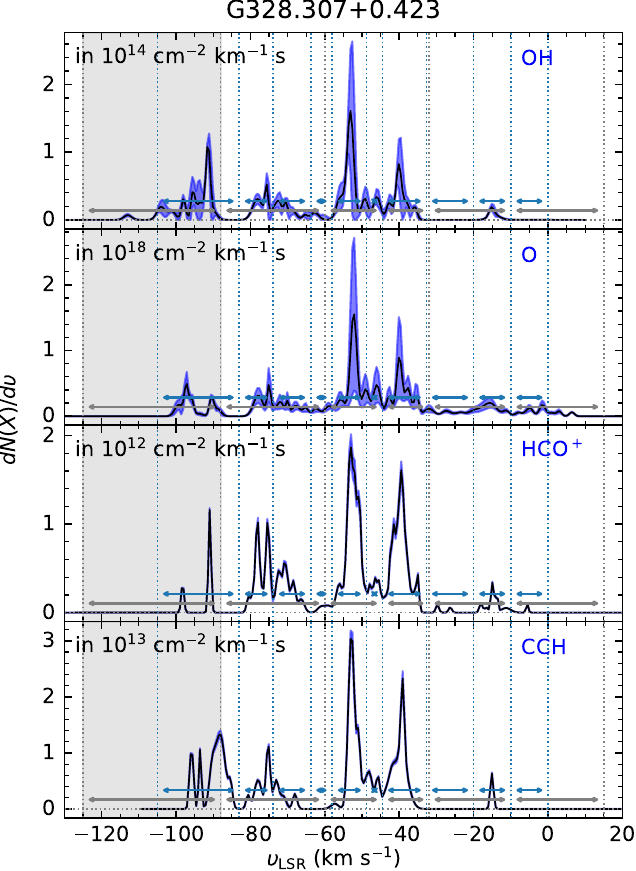}
    \includegraphics[width=0.195\textwidth]{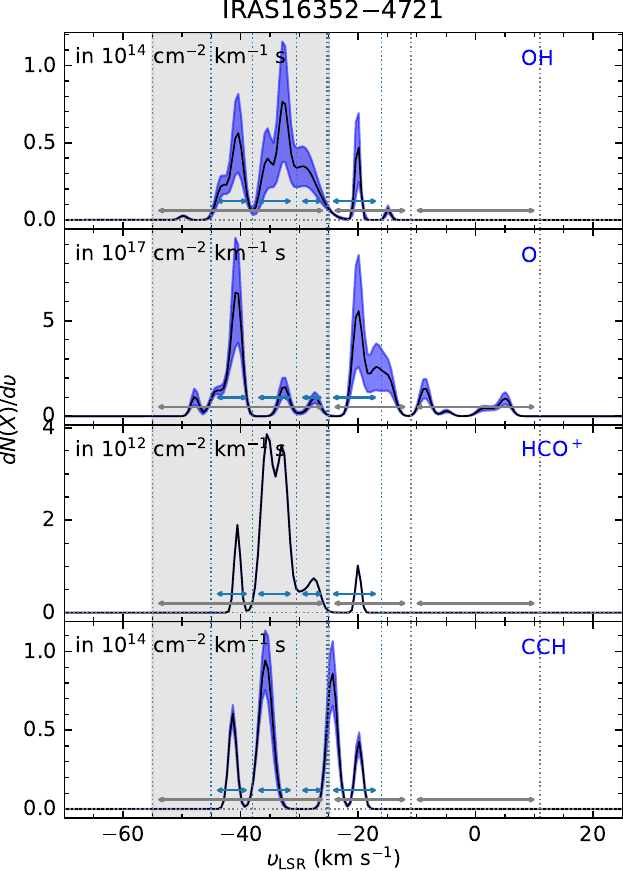}
    \includegraphics[width=0.195\textwidth]{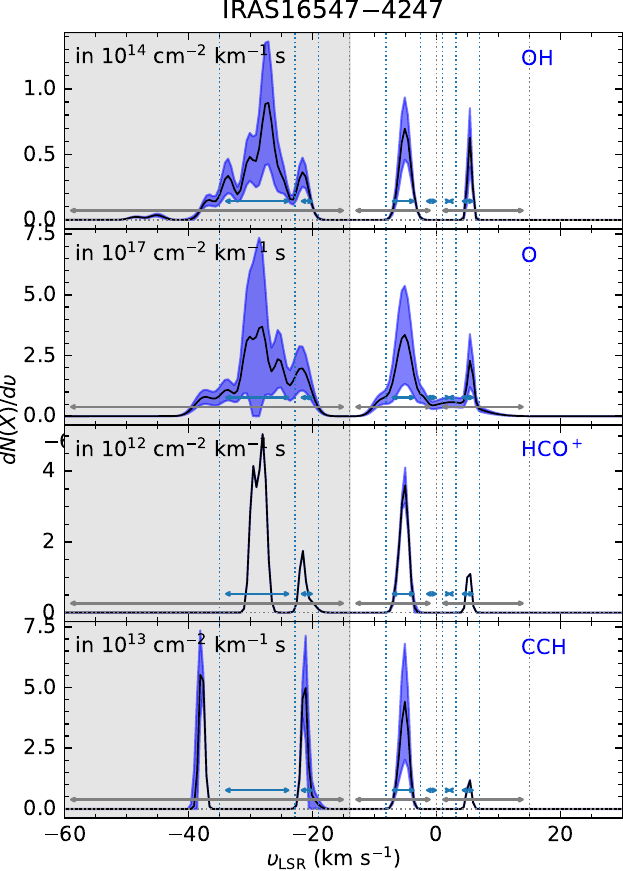}
    \includegraphics[width=0.198\textwidth]{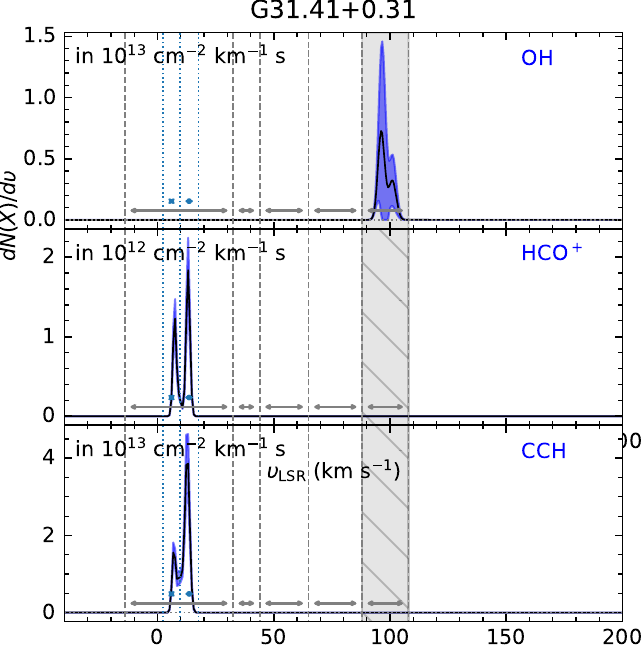}
    \includegraphics[width=0.195\textwidth]{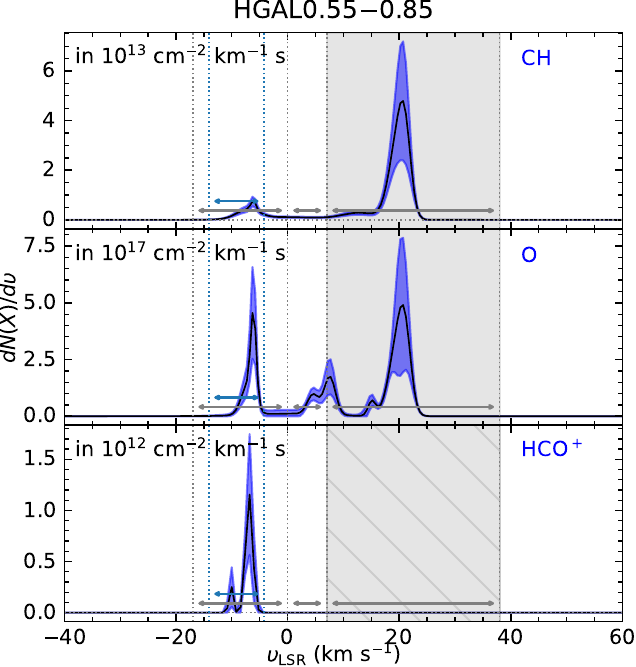}
    \includegraphics[width=0.195\textwidth]{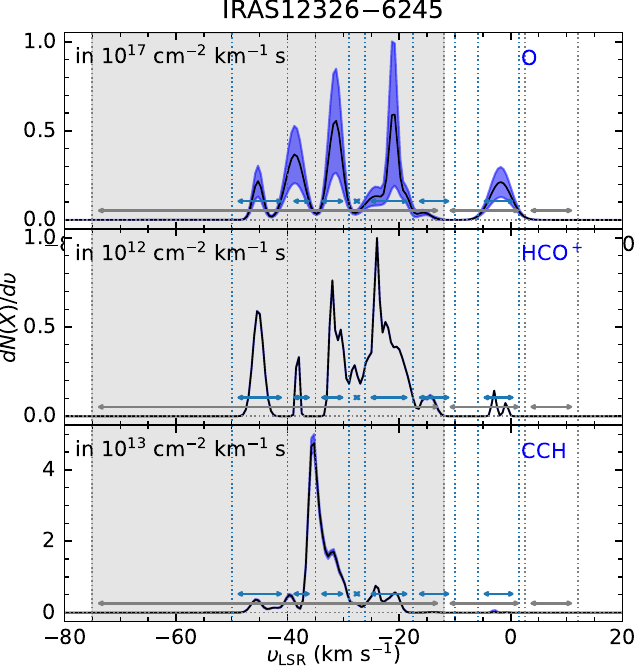}
    \includegraphics[width=0.195\textwidth]{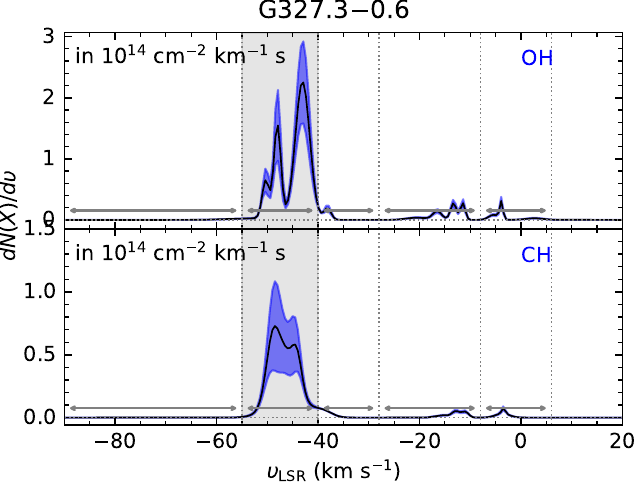}
    \includegraphics[width=0.195\textwidth]{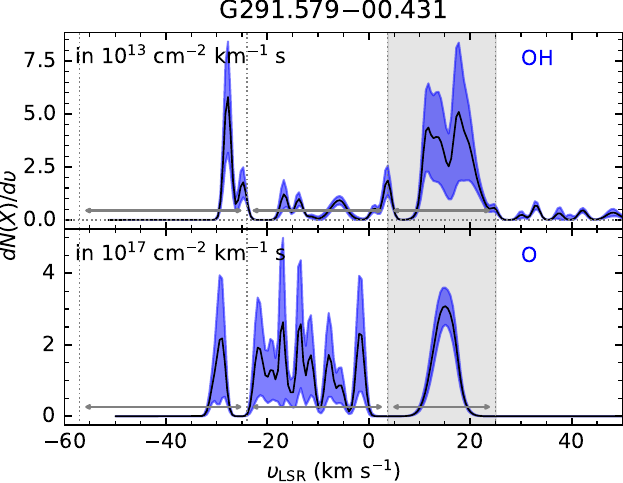}
    \includegraphics[width=0.195\textwidth]{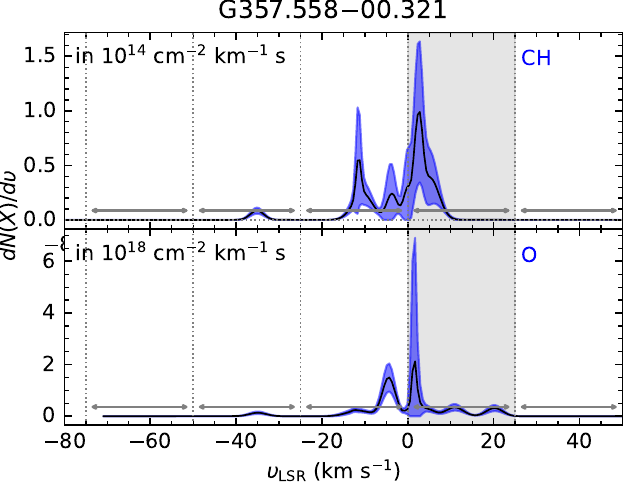}
    \includegraphics[width=0.195\textwidth]{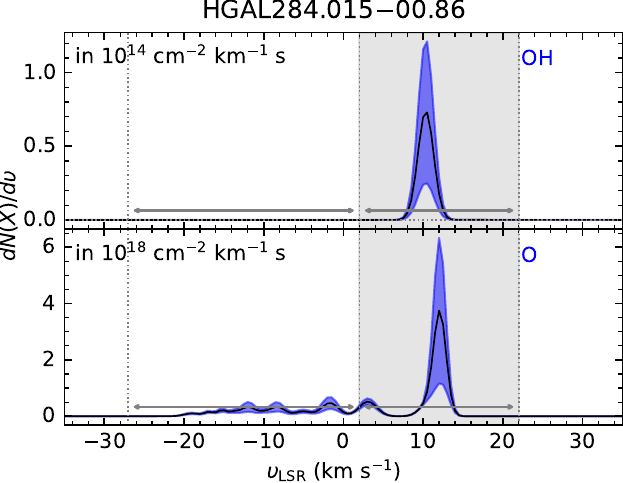}
    \includegraphics[width=0.195\textwidth]{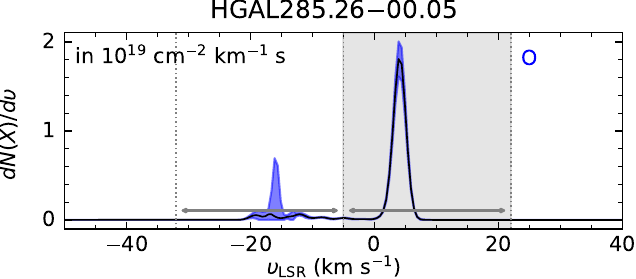}
    \caption{Channel-wise column density spectra ($Nd/d\varv$) as a function of velocity. Same as Fig.\,\ref{fig:Ncol_spec}}
    \label{appendix:Ncol}
\end{figure*}

\begin{table*}[h!]
\begin{center}
\tiny
\caption{Derived column densities over specific velocity intervals (VI(A)) and molecular gas fractions.}
\label{tab:column_density_tb1}
\begin{tabular}{lcccccccccc}
\hline\hline
Source name & VI(A) & $R_{\rm GAL}$& $N$(CH) & $N$(OH) & $N$(\ce{HCO+}) & $N$(CCH) & $N$(O) & $N$(\ce{H2})$^\ddagger$ & $N$(H~{\sc i})$^\dagger$ & $f^{N}_{\ce{H2}}$$^\ddagger$ \\
 & $\upsilon_{\rm min}$, $\upsilon_{\rm max}$ & $R_{\rm g, min}$, $R_{\rm g, max}$ & $\times10^{13}$ & $\times10^{14}$ & $\times10^{12}$ & $\times10^{13}$ & $\times10^{18}$ & $\times10^{21}$ & $\times10^{22}$ & \\
       & [\kms] &  [kpc] & [cm$^{-2}$] & [cm$^{-2}$] & [cm$^{-2}$]  & [cm$^{-2}$] & [cm$^{-2}$] & [cm$^{-2}$] & [cm$^{-2}$] &  \\ 
\hline 
HGAL284.015$-$00.86 & $-27$, $+$2 & 7.3, 8.2 & $\cdots$ & $\cdots$ & $\cdots$  & $\cdots$ & 4.70$^{+1.85}_{-1.85}$ & $\cdots$ & 3.63$^{+2.20}_{-1.11}$ & $\cdots$ \\ 
\hline
HGAL285.26$-$00.05  & $-32$, $-5$ & 7.1, 8.0 & $\cdots$ & $\cdots$  & $\cdots$ & $\cdots$ & 5.57$^{+3.44}_{-3.44}$ & $\cdots$ & 3.33$^{+1.86}_{-0.99}$ & $\cdots$ \\ 
\hline
G291.579$-$00.431   & $-57$, $-24$ & 6.5, 7.3 & $\cdots$ & 1.33$^{+0.59}_{-0.59}$ & $\cdots$ & $\cdots$ & 0.56$^{+0.40}_{-0.40}$ & 1.08$^{+0.48}_{-0.48}$ $^a$ & 0.88$^{+0.90}_{-0.40}$  & 0.099 \\ 
                    & $-24$, $+$3.7 & 7.3, 8.3 &$\cdots$ & 1.10$^{+0.41}_{-0.41}$  & $\cdots$ & $\cdots$ & 2.88$^{+1.98}_{-1.98}$ & 0.89$_{-0.34}^{+0.34}$ $^a$ & 1.73$^{+0.97}_{-0.43}$ & 0.047 \\ 
\hline
IRAS~12326-6245 & $-12$, $+$2.5 & 7.7, 8.2 & $\cdots$ & $\cdots$ & 0.21$^{+0.01}_{-0.01}$ & 0.03$^{+0.02}_{-0.02}$ & 1.01$^{+0.37}_{-0.37}$ & $\cdots$ & 0.21$^{+0.03}_{-0.01}$ & $\cdots$ \\ 
                & $+$2.5, $+$12 & 8.2, 8.7 & $\cdots$ & $\cdots$ & $\cdots$ & $\cdots$ &  0.01$^{+0.003}_{-0.003}$ & $\cdots$ & 0.56$^{+0.09}_{-0.03}$ & $\cdots$ \\ 
                & $+$12, $+$65 & 8.7, 12.0 & $\cdots$ & $\cdots$ & $\cdots$ & $\cdots$ & $\cdots$ & $\cdots$ & 0.25$^{+0.02}_{-0.01}$ & $\cdots$ \\ 
\hline
G327.3$-$00.60 	& $-90$, $-55$	& 4.8, 5.7 & 0.07$^{+0.02}_{-0.02}$ & 0.08$^{+0.01}_{-0.01}$  & $\cdots$ & $\cdots$ & $\cdots$ & 0.02$^{+0.01}_{-0.01}$ & 0.63$^{+0.01}_{-0.01}$  & 0.003 \\ 
                & $-40$, $-28$ & 6.2, 6.7 & 1.81$^{+0.05}_{-0.05}$ & 0.49$^{+0.09}_{-0.09}$  & $\cdots$ & $\cdots$ & $\cdots$ & 0.15$^{+0.01}_{-0.01}$  & 0.15$^{+0.01}_{-0.01}$ & 0.083 \\
                & $-28$, $-8$ & 6.7, 7.7 & 2.36$^{+0.39}_{-0.39}$ & 1.23$^{+0.26}_{-0.26}$  & $\cdots$ & $\cdots$ & $\cdots$ &  0.67$^{+0.11}_{-0.11}$ & 0.22$^{+0.01}_{-0.01}$ & 0.189 \\ 
                & $-8$, $+$6 & 7.7, 8.5 & 1.49$^{+0.25}_{-0.25}$ & 0.56$^{+0.11}_{-0.11}$  & $\cdots$ & $\cdots$ & $\cdots$ & 0.42$^{+0.07}_{-0.07}$ & 0.21$^{+0.01}_{-0.01}$ & 0.143 \\ 

\hline
\end{tabular}
\end{center}
\tablefoot{The full table is available at the CDS via anonymous ftp to cdsarc.cds.unistra.fr (130.79.128.5). This table only presents a fraction of the entire column density table. ($^*$) The attribution of velocity intervals is based mainly on the HI column density profiles and supplementary \ce{OH+} (Jacob et al. in prep) and OH column density profiles from this work. The Galactocentric distance for each velocity range is determined by $R_{\rm GAL}=R_0\frac{\Theta(R_{\rm GAL}){\rm sin}(l){\rm cos}(b)}{\upsilon_{\rm LSR} + \Theta_0{\rm sin}(l){\rm cos}(b)}$, with $R_0=8.15$\,kpc and $\Theta_0 = 236$\,\kms\ \citep{Reid2014}, and $\Theta(R_{\rm GAL})$ becomes $\Theta_0$ with the assumption of a flat Galactic rotation curve. ($\ddagger$) The $N$(\ce{H2}) and molecular gas fraction ($f^{N}_{\ce{H2}}=\frac{2N(\ce{H2})}{N({\rm HI})+2N(\ce{H2})}$) are determined by adopting the abundance of CH relative to \ce{H2}, $N$(CH)/$N$(\ce{H2}) $=3.5\times10^{-8}$ \citep{Sheffer2008}. For the case ($^a$) in the absence of CH data, we used the measured column density ratio of $N$(OH)/$N$(CH)~$=$~3.29 to convert to $N$(\ce{H2}). ($\dagger$) The atomic hydrogen column densities are obtained from Rugel et al. (in prep.). } 
\end{table*}
\newpage
\section{Chemical networks of the observed species}

\begin{figure*}[h!]
\centering
\includegraphics[width=0.6\textwidth]{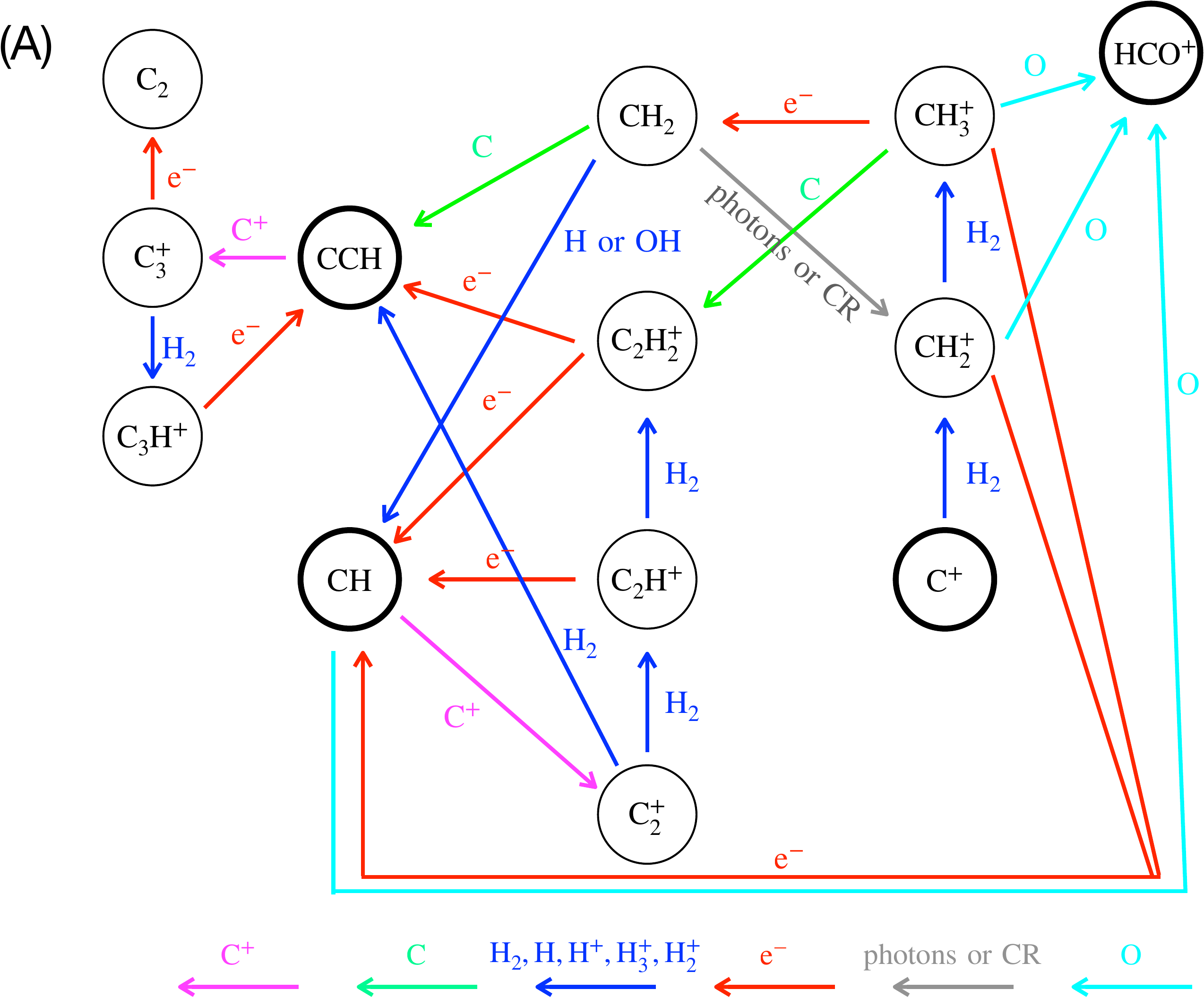}\vspace{0.4cm}
    \includegraphics[width=0.6\textwidth]{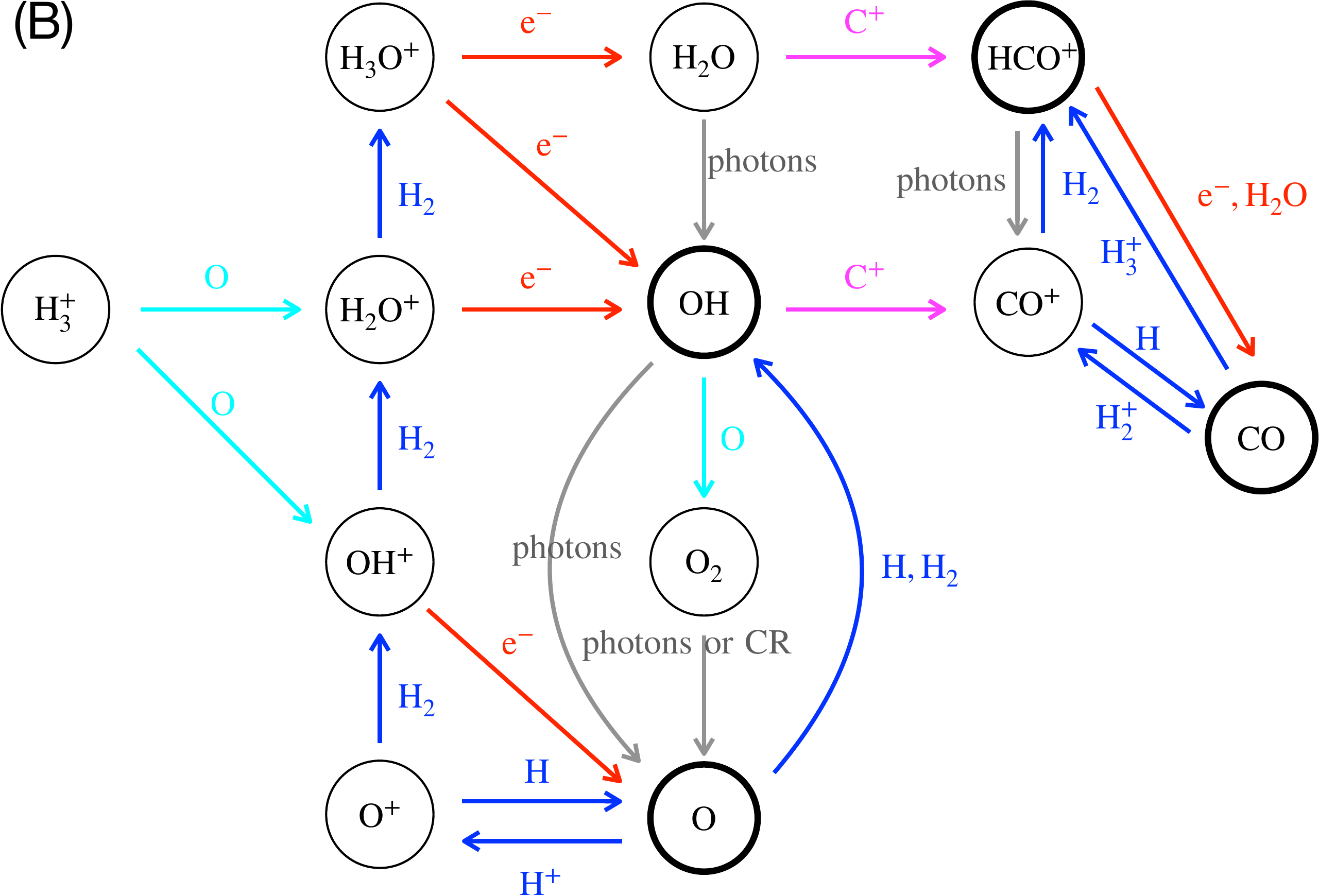}
    \caption{Sketches of chemical networks for carbon-bearing (A panel) and oxygen-bearing species (B panel) relevant to this work. Both networks are connected through \ce{HCO+} and CO \citep{van_Dishoeck1986_chemical_networks}. These sketches are simplified chemical networks that demonstrate the major reactions among the species studied in this work, based on the literature \citep[][and references therein]{Sternberg1995_chemical_networks,Godard2009_TDR,Godard2014_TDR,Balashev2021_OH_diffuse}. The colors and labels of the arrows indicate the reaction partners. We note that these networks represent generally possible chemical reactions, rather than those requiring specific physical conditions.}
    \label{fig:chemical_networks}
\end{figure*}
\newpage

\section{Column density ratios versus Galactocentric distance and molecular gas fraction}

\begin{figure}[h!]
    \centering
    \includegraphics[width=0.38\textwidth]{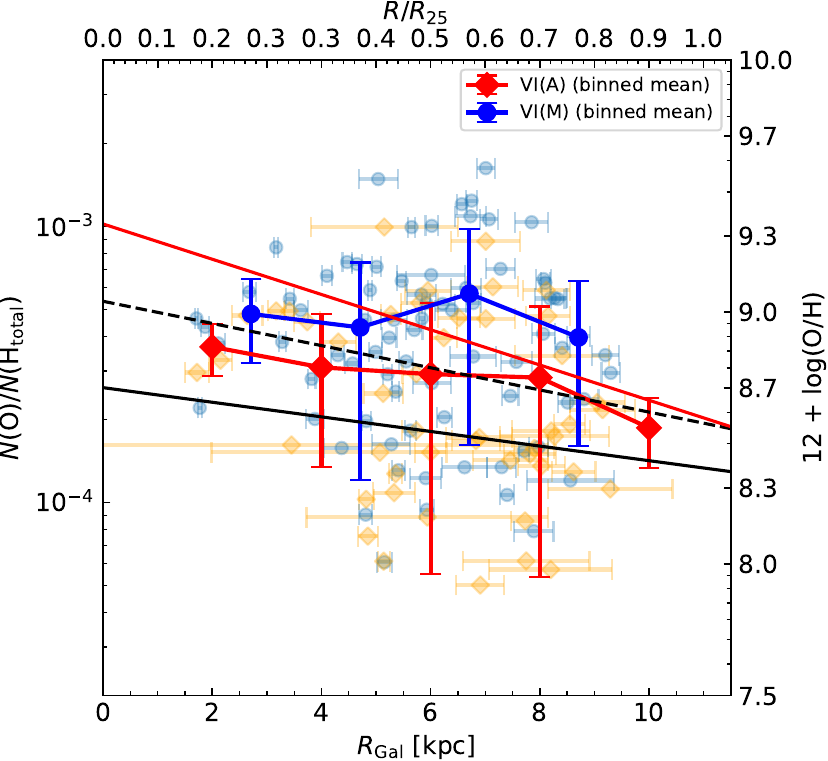}
    \hspace{1cm}
    \includegraphics[width=0.38\linewidth]{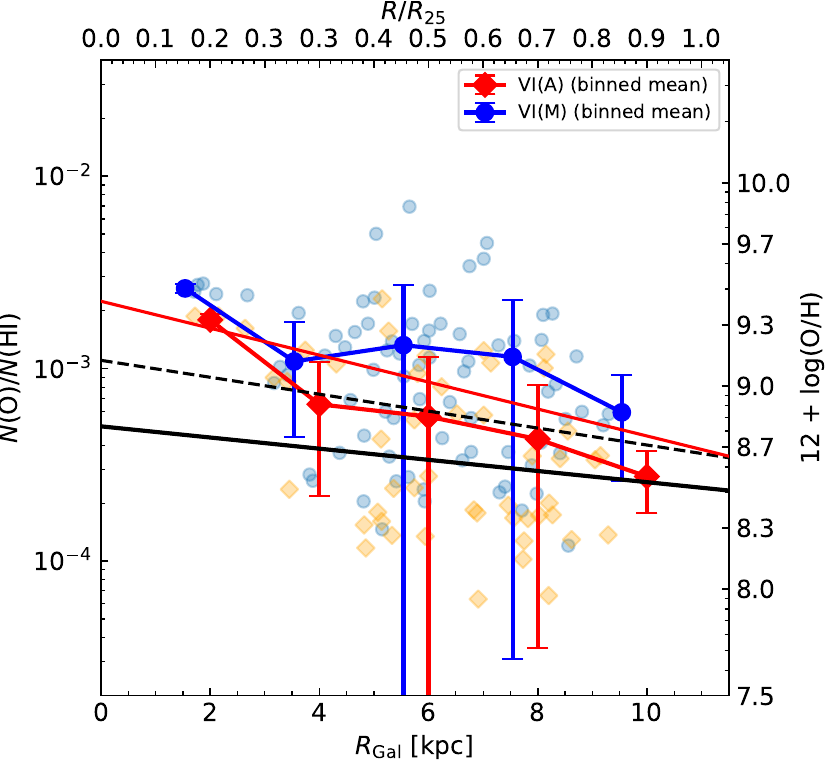}
        \caption{(\textit{Left}) Column density ratio (left y-axis) of neutral oxygen to total hydrogen versus Galactocentric distance (lower x-axis). (\textit{Right}) Column density ratio of neutral atomic oxygen and neutral atomic hydrogen across the Galactocentric distance.  The orange diamond and blue circle markers represent the VI(A) and VI(M) samples, respectively. The error bars in $R_{\rm GAL}$ are the ranges of Galactocentric distances. The red squares and blue circles depict the averaged ratios for each sample group over 2 kpc bins, with error bars indicating the standard deviations. The black solid and dashed lines illustrate the O/H relationship measured with ionized oxygen (\ce{O+} and \ce{O++}) and hydrogen (\ce{H+}) recombination lines in Galactic H{\sc ii} regions taken from \cite{Esteban2005_HII_metalicity_MW} and \cite{Esteban2017_O_H_HII}, respectively. The red solid line is the O/H relationship derived using the atmospheric absorption spectral lines toward young B-type stars \citep{Smartt2001_O_H_inner_MW}. On the right y-axis, the column density ratio is expressed as oxygen abundance, while the upper x-axis shows the Galactocentric distance in terms of $R/R_{\rm 25}$, where $R_{\rm 25} = 11.5$ kpc \citep{Esteban2005_HII_metalicity_MW,Esteban2017_O_H_HII}.}
    \label{fig:Rgal_No_Nhtotal}
\end{figure}

To investigate whether the variation in the neutral oxygen abundance within the Galactic diffuse ISM is linked to the Galactic metallicity gradient, we examined the ratios of neutral oxygen to total hydrogen column densities as a function of Galactocentric radius. As shown in the left panel of Fig.~\ref{fig:Rgal_No_Nhtotal}, both velocity interval samples exhibit broadly similar distributions. The VI(M) sample shows slightly higher ratios than the VI(A) sample, as reflected in the 2\,kpc-binned averages (red and blue curves). However, considering the uncertainties presented in Fig.~\ref{fig:No_Nhi_Nh2}, these differences are not statistically significant. Overall, $N$(O)/$N$(H$_{\rm total}$) from this work displays substantial scatter. In addition, the averaged ratios (blue circles and red diamonds) are broadly consistent with the O/H values derived from recombination lines in Galactic H~{\sc ii} regions \citep{Esteban2005_HII_metalicity_MW, Esteban2017_O_H_HII} (dashed and solid black lines), as well as with those measured from stellar atmospheric absorption lines toward B-type stars within the inner 5\,kpc of the Galactic disk \citep{Smartt2001_O_H_inner_MW} (solid red line).

The large dispersions in our data points may result from line-of-sight integration through multiple ISM phases with distinct metallicities, particularly between 4 and 8\,kpc. Because absorption-line measurements are integrated over velocity intervals defined by line profiles, they inevitably sample gas components with different physical conditions. In addition, since the plotted $R_{\rm GAL}$ values represent mean distances corresponding to each velocity interval, the derived abundances effectively average over Galactocentric ranges of $1-4$\,kpc (corresponding to the x-error bars), likely encompassing several spiral arms such as the Sagittarius–Carina, Norma–Outer, and Scutum–Centaurus arms as well as inter-arms. In alignment with the aforementioned findings, the radio recombination line study conducted by \cite{Balser2011_O_H_MW} revealed azimuthal variations in the O/H gradient ranging from $-0.03$ to $-0.07$\,dex\,kpc$^{-1}$, implying that the metallicity distribution within the Galactic disk exhibits deviations from strict axisymmetry. The observed azimuthal variations in the Galactic O/H gradient from the study by \cite{Balser2011_O_H_MW} may therefore account, at least in part, for the significant scatter seen in our data.

Alternatively, the observed scatter could reflect intrinsic variations in O/H that depend on the gas phase being traced (see details in Sect.\,\ref{sec:abundance_fraction})—whether ionized or neutral. Toward the external galaxy M83, \citet{Hernandez2021_externalgalaxies_metalicity} reported that the O/H ratios derived from neutral gas exhibit a flat radial trend, whereas those from ionized gas and stellar measurements show clear gradients with Galactocentric radius. According to the results from \cite{Hernandez2021_externalgalaxies_metalicity}, in the nuclear region of M83, the metallicity of the neutral gas exceeds that of the ionized gas. However, our results toward the Galactic ISM do not show a significant enhancement in oxygen abundance toward the Galactic center (around $R_{\rm GAL}\sim2\,$kpc) compared with those found toward M83 \citep{Hernandez2021_externalgalaxies_metalicity}. This discrepancy may arise from differences in the methods used to estimate \ce{H2}. Our analysis is based on CH and OH absorption lines, which also trace CO-dark \ce{H2}, whereas extragalactic studies typically rely on CO emission as a proxy for molecular hydrogen. Nevertheless, both M83 and the inner MW exhibit high \ce{H2} column densities and low neutral atomic hydrogen fractions, consistent with an overall increase in oxygen abundance relative to atomic hydrogen toward the Galactic center (right panel).

\begin{figure*}[h!]
\centering 
    \includegraphics[width=0.38\textwidth]{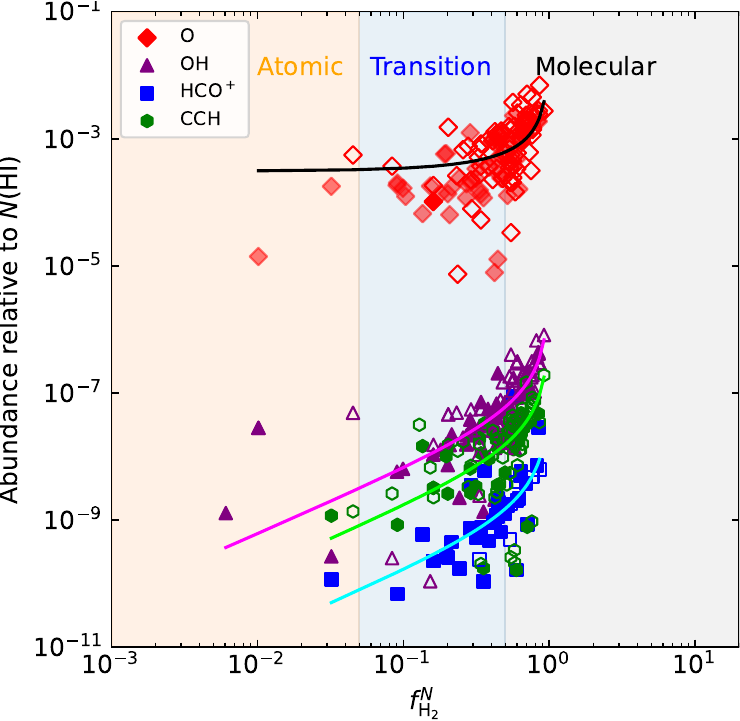}
    \caption{Same as Fig.\,\ref{fig:x_fmol} but O, OH, \ce{HCO+}, and CCH abundances relative to $N$(HI) (left panel) and to $N$(\ce{H2}) (right panel) versus $f_{\ce{H2}}^N$. The colored dotted lines are median abundance values of OH (purple), \ce{HCO+} (cyan), and CCH (lime). }
    \label{appendix:abundances}
\end{figure*}
Figure\,\ref{appendix:abundances} displays relative abundances of O, OH, \ce{HCO+}, and CCH to neutral atomic hydrogen (H~{\sc i}; left panel) and molecular hydrogen (\ce{H2}; right panel) as a function of molecular gas fraction ($f_{\ce{H2}}^N$).
\end{appendix}

%
%

\end{document}